\documentclass[letterpaper,titlepage,11pt]{article}

\pdfoutput=1

\usepackage{amssymb,amsmath,amsfonts}
\usepackage{epsfig}
\usepackage{ccaption}
\usepackage{graphicx}

\usepackage[
      colorlinks=true,
      linkcolor=blue,
      urlcolor=blue,
      filecolor=black,
      citecolor=red,
      pdfstartview=FitV,
      pdftitle={},
        pdfauthor={Roberto Emparan, Ryotaku Suzuki, Kentaro Tanabe},
        pdfsubject={},
        pdfkeywords={},
        pdfpagemode=None,
        bookmarksopen=true
      ]{hyperref}

\def\clock{{\count0=\time
           \divide\count0 60
           \ifnum\count0<10 0\fi\the\count0
           \multiply\count0 -60 \advance\count0 \time
           :\ifnum\count0<10 0\fi \the\count0
         }}
\newcommand{\timestamp}{{\small\vbox{\hbox{\tt\jobname.tex}
\hbox{\the\day/\the\month/\the\year, \clock}}}}

\newcommand{\ie}{{\it i.e.,\,}}
\newcommand{\eg}{{\it e.g.,\,}}

\newcommand{\lp}{\left(}
\newcommand{\rp}{\right)}

\newcommand{\beq}{\begin{equation}}
\newcommand{\eeq}{\end{equation}}
\newcommand{\bea}{\begin{eqnarray}}
\newcommand{\eea}{\end{eqnarray}}
\newcommand{\beqa}{\begin{eqnarray}}
\newcommand{\eeqa}{\end{eqnarray}}

\newcommand{\sR}{\mathsf{R}}

\numberwithin{equation}{section}

\begin{document}

\begin{titlepage}
\leftline{}
\vskip 1cm
\centerline{\LARGE \bf The large $D$ limit of General Relativity}
\vskip 1.6cm
\centerline{\bf Roberto Emparan$^{a,b}$, Ryotaku Suzuki$^{c}$, Kentaro Tanabe$^{b}$}
\vskip 0.5cm
\centerline{\sl $^{a}$Instituci\'o Catalana de Recerca i Estudis
Avan\c cats (ICREA)}
\centerline{\sl Passeig Llu\'{\i}s Companys 23, E-08010 Barcelona, Spain}
\smallskip
\centerline{\sl $^{b}$Departament de F{\'\i}sica Fonamental, Institut de
Ci\`encies del Cosmos,}
\centerline{\sl  Universitat de
Barcelona, Mart\'{\i} i Franqu\`es 1, E-08028 Barcelona, Spain}
\smallskip
\centerline{\sl $^{c}$Department of Physics, Kyoto University, Kyoto 606-8502, Japan}
\vskip 0.5cm
\centerline{\small\tt emparan@ub.edu,\, ryotaku@tap.scphys.kyoto-u.ac.jp,\, ktanabe@ffn.ub.es}

\vskip 1.6cm
\centerline{\bf Abstract} \vskip 0.2cm \noindent
General Relativity simplifies dramatically in the limit that the number of spacetime dimensions $D$ is infinite: it reduces to a theory of non-interacting particles, of finite radius but vanishingly small cross sections, which do not emit nor absorb radiation of any finite frequency. Non-trivial black hole dynamics occurs at length scales that are $1/D$ times smaller than the horizon radius, and at frequencies $D$ times larger than the inverse of this radius. This separation of scales at large $D$, which is due to the large gradient of the gravitational potential near the horizon, allows an effective theory of black hole dynamics. We develop to leading order in $1/D$ this effective description for massless scalar fields and compute analytically the scalar absorption probability. We solve to next-to-next-to-leading order the black brane instability, with very accurate results that improve on previous approximations with other methods. These examples demonstrate that problems that can be formulated in an arbitrary number of dimensions may be tractable in analytic form, and very efficiently so, in the large $D$ expansion. 

\end{titlepage}
\pagestyle{empty}
\small
\normalsize
\newpage
\pagestyle{plain}
\setcounter{page}{1}

\addtocontents{toc}{\protect\setcounter{tocdepth}{2}}

\tableofcontents
\newpage

\section{Introduction}

The fascination power of General Relativity stems largely from the wealth of physical phenomena that are encoded in equations as simple as
\beq\label{Rmunu}
R_{\mu\nu}=0\,.
\eeq
Naturally, this conceptual simplicity unfolds its rich dynamics at the cost of technical complexity. It is very difficult to find closed exact solutions to these coupled, non linear, partial differential equations, for almost any phenomenon of interest unless a substantial degree of symmetry is present. 

Physical theories often contain parameters that can be varied in such a way that the theories remain well defined. A fruitful strategy is to focus on regions of the parameter space, usually close to its origin or boundaries, where the theories simplify. Einstein's theory in vacuum, \eqref{Rmunu}, appears to have only one natural parameter: the number $D$ of spacetime dimensions. We will argue that in the limit $D\to\infty$, General Relativity simplifies dramatically, its dynamics becoming trivial at all non-zero length scales away from the horizons of black holes. This is a strong motivation for the study of the theory in an expansion in $1/D$.
 
While it seems unlikely that our universe be infinite-dimensional, the study of General Relativity around this limit can be useful, both to gain a better understanding of the theory and as an approximation scheme for calculations in less unrealistic cases, say $D=4$ or $D=10,11$. 
We have learned in recent years that new features appear in the spectrum of black hole solutions as $D$ grows beyond four \cite{hidbook}. Still, it may not be unreasonable to expect that some properties remain more robust as $D$ is increased --- to begin with, the theory \eqref{Rmunu} does not have black holes nor a dynamical graviton when $D\leq 3$, and always has them for any $D\geq 4$. The question of to what extent an expansion in $1/D$ is a good qualitative guide to moderate $D$, and if so, how accurate it is, probably depends on the specific problem under consideration. In this article this concern will remain mostly in the background, and instead we will focus on understanding the main properties of the limit and on how to organize calculations in the $1/D$ expansion. Nevertheless, one of our examples shows that these techniques can give very accurate results even at relatively low values of $D$.

Early studies of gravity in the large $D$ limit analyzed the quantum theory and the properties of its Feynman diagrams \cite{Strominger:1981jg,BjerrumBohr:2003zd} (see also \cite{Hamber:2005vc}). The main motivation is the possible analogy with the large $N$ limit of $SU(N)$ gauge theories --- indeed, the local Lorentz group $SO(D-1,1)$ is the gauge group of gravity. 
The large $N$ limit of Yang-Mills theories is useful because, although the number of gluons grows unbounded, they arrange themselves into worldsheets of strings (propagating in more dimensions). One might hope that a similar miracle could occur also for large $D$ gravity. However, $D$ appears not only in the number of graviton polarizations but also, more troublingly, in phase space integrals. Actually, taking $D$ to be large seems a bad idea for a  quantum field theory, since the ultraviolet behavior generically worsens. This can be alleviated by focusing on Kaluza-Klein truncations of the spectrum  \cite{Strominger:1981jg,BjerrumBohr:2003zd,Canfora:2009dx}, which retain the growing number of polarizations but make the short-distance behavior essentially four-dimensional, with the usual divergence problems in the gravitational sector. We shall not pursue any of these approaches.

Instead, we study mostly the classical theory of eq.~\eqref{Rmunu}, which is well defined in any $D$. 
After all, many quantum properties of gravity are dominated by classical effects such as black hole formation. One speculation is that the illnesses of quantum gravity may be absent outside the horizons when $D\to\infty$ and might remain under some control close to the horizon in the $1/D$ expansion. At any rate we will see that, even if its quantum version happens to be badly behaved, the large $D$ limit of the classical theory is a useful one.

The study of classical gravity in arbitrary $D\geq 4$ has gained momentum over the years, and it has often seemed natural to examine specific results in the limit $D\to\infty$. However, very rarely has the large $D$ expansion been pursued as a subject in its own right. A notable exception are refs.~\cite{Kol:2004pn,Asnin:2007rw}, which are closest in spirit and techniques to our approach. Nevertheless, their context was restricted to a particular phenomenon (the Euclidean zero mode of the Schwarzschild solutions) and a bigger framework was not developed.
Some general observations about black holes and black branes at large $D$ are made in \cite{Caldarelli:2008mv,Camps:2010br} emphasizing slightly different features than here. Although there are many studies of gravitational phenomena in arbitrary $D$, those that make explicit reference to the large $D$ limit are, as far as we are aware, relatively  few, 
\eg~\cite{Soda:1993xc}--\cite{Caldarelli:2012hy}.

In summary: a systematic approach to the large $D$ limit of classical General Relativity has been lacking so far. We aim to provide some basic entries to the concepts and techniques of this subject. 

A main element of our approach is that, in contrast to the studies inspired by large $N$ gauge theories, we will not focus on perturbations around a Minkowski background but rather on non-perturbative objects in the theory, namely its black holes. We argue that in the limit $D\to\infty$ black holes behave in many respects like non-interacting particles when probed at the scales that are natural to observers away from them. They do not attract each other and, even if their radius remains finite, their collision cross sections vanish. They reflect perfectly all radiation of any finite frequency. 
Black branes do behave like if made of dust, with no tension to hold them together. 

The simplicity of this limit makes it a good starting point for a perturbative expansion in $1/D$. We will see that in the limit $D\to\infty$ the gravitational field vanishes outside the horizon at $r=r_0$. This field is strongly localized close to the horizon in the region
\beq\label{nearhor}
r-r_0\lesssim\frac{r_0}{D}\,.
\eeq
Crucially, a new scale appears owing to the very steep gradient, $\sim D/r_0$, of the gravitational potential near the horizon. 
This separation between scales $r_0/D\ll r_0$ allows to develop an effective theory of black hole dynamics. Fields outside the black hole propagate in an effectively flat spacetime, subject to certain boundary conditions very near the horizons, which replace the black holes. Technically, this is a problem of matched asymptotic expansions. Conceptually, it is an effective theory in which the degrees of freedom for the black hole, at the frequency scale $D/r_0$, are integrated out and replaced by boundary conditions on large-distance fields. What is peculiar to this effective theory is that the notion of short-distance degrees of freedom is the result of having large $D$, instead of the more conventional idea of considering wavelengths much larger than the horizon radius. This results in a much larger range of applicability of the effective theory.

We begin in the next section motivating the `non-interacting particle' picture of the limit $D\to\infty$ through an extensive study of known black hole solutions. In section~\ref{sec:sphinf} we introduce the notion of the `sphere of influence' of the black hole. In section~\ref{sec:radn} we discuss how and when classical gravitational radiation can be emitted through black hole interactions at $D\to\infty$. In section~\ref{sec:quantum} we make some comments about quantum effects and Hawking radiation in black holes at large $D$. In section~\ref{sec:scalar} we introduce the $1/D$ expansion in the study of the propagation of massless scalars in the black hole background. We solve the theory in the region near the horizon and find the effective boundary conditions for outside fields. We obtain a compact analytic expression for the scalar absorption probability that is valid over a very wide range of frequencies. In section~\ref{sec:GL} we solve the spectrum of unstable perturbations of black branes in closed analytic form to next-to-next-to-leading order at large $D$.  The results are very accurate for all but the two lowest dimensions. We conclude in section~\ref{sec:concl}.

\section{Large $D$ limit of black holes}
\label{sec:largeD}

We take the point of view that General Relativity in vacuum, the theory of \eqref{Rmunu}, is essentially a theory of black holes that interact via the gravitational field between them, and which can emit and absorb gravitational waves through these interactions. We can also include as objects of study the singular plane wave solutions that appear in the infinite boost limit of black holes, but in general we exclude other nakedly singular solutions. 

We begin with the most basic solution, the Schwarzschild-Tangherlini spacetime 
\beq\label{schwd}
ds^2=-f dt^2+f^{-1}dr^2+r^2 d\Omega_{D-2}\,,
\eeq
\beq
f(r)=1-\lp\frac{r_0}{r}\rp^{D-3}\,,
\eeq
with a horizon of radius $r_0$ \cite{Tangherlini:1963bw}. 
When we take the limit $D\to\infty$ we have to decide how the magnitudes in the problem scale with $D$, and in particular which quantities are kept fixed, \ie\ do not scale with $D$. This amounts to deciding the regime of physics one focuses on. In particular, if a certain dimensionful quantity is kept fixed, then we are selecting physics at the scale of that quantity. 

In the present case, if we keep $r_0$ fixed as $D\to\infty$ then the metric remains finite at all $r>r_0$. In this manner we expect to capture the physics of wavelengths parametrically comparable (in the parameter $D$) to $r_0$, and frequencies comparable to $1/r_0$, in the region outside the horizon. Note that by setting $r_0$ as the relevant scale, all dimensionful quantities can be appropriately rendered dimensionless by dividing them by a power of $r_0$, or equivalently by setting $r_0=1$. We find that the discussion is often clearer if we keep $r_0$ explicit. 

While the choice of $r_0$ as the scale to be fixed appears rather natural, a main point of this article is that scales much smaller than $r_0$ are also present when $D$ is a large parameter.

\subsection{`Smallness' of the horizon and a hierarchy of scales}
\label{subsec:small}

Some of the properties of black holes at large $D$ are not due to spacetime curvature but rather follow from elementary flat-space geometry (see appendix~\ref{app:elemgeo}). In particular, when $D$ grows large, the area of the unit-radius sphere $S^{D-2}$,
\beq\label{OmD2}
\Omega_{D-2}=\frac{2\pi^{(D-1)/2}}{\Gamma\lp\frac{D-1}{2}\rp}\,,
\eeq
vanishes as
\beqa\label{tozero}
\Omega_{D-2}\to \frac{D}{\sqrt{2}\pi}\lp \frac{2\pi e}{D}\rp^{D/2}\to 0\,.
\eeqa
We may then say that these spheres becomes increasingly small at very large $D$, shrinking at a rate $\sim D^{-D/2}$. In particular, if we shoot a projectile at this unit-radius sphere, the impact cross section vanishes when $D\to \infty$, even if the projectile will hit the target whenever the impact parameter is $< 1$.\footnote{The actual cross section of the unit $S^{D-2}$ is $\sigma=\Omega_{D-3}/(D-2)$, whose dominant large $D$ behavior is inversely semifactorial like that of $\Omega_{D-2}$.} This geometric effect will be present in all calculations of total cross sections at $D\to\infty$. The strong semifactorial suppression $\sim D^{-D/2}$ may sometimes hide effects that one is interested in, and in these cases it may be convenient to eliminate it by, \eg\ considering ratios of appropriate magnitudes.

The unconventional property of the limit $D\to\infty$ that spheres can have {\it finite radius but zero area} implies that the concept of ``black hole size'' is ambiguous when $D$ is large. To a sphere of radius $r_0$, such as the horizon of \eqref{schwd}, we can associate two length scales: the radius $r_0$ itself, and a much smaller ``area length"\footnote{This must not be confused with the ``area radius" defined as $(A_H/\Omega_{D-2})^{1/(D-2)}$, which is indeed $r_0$.}
\beq\label{larea}
\ell_A \sim A_{H}^{1/(D-2)}\sim \frac{r_0}{\sqrt{D}}\,.
\eeq 

There is another scale that will be much more relevant in black hole physics and which is independent of the previous flat-space, non-gravitational effect. It is due to the large radial gradient of the gravitational potential near the horizon,
\beq
\left. \partial_r f\right|_{r_0} \to \frac{D}{r_0}\,.
\eeq
For instance, the surface gravity on the horizon is
\beq
\kappa=\frac{D-3}{2 r_0}\,,
\eeq
which becomes much larger than the scale $1/r_0$ when $D\to\infty$. Thus a length-scale arises associated to $\kappa$,
\beqa\label{elk}
\ell_\kappa=\kappa^{-1}
\sim\frac{r_0}{D}
\eeqa
which is parametrically smaller than the horizon radius $r_0$. 
The intrinsic curvature gives essentially this same length: the Kretschmann scalar in the geometry \eqref{schwd} is
\beq\label{krets}
K=R_{\mu\nu\sigma\rho}R^{\mu\nu\sigma\rho}=\frac{(D-1)(D-2)^2(D-3)}{r^4}\lp\frac{r_0}{r}\rp^{2(D-3)}
\eeq
so the characteristic curvature length of the horizon is
\beq\label{lK}
\lp K(r_0)\rp^{-1/4}\to\frac{r_0}{D}\sim \ell_\kappa\,.
\eeq

Of the two small scales that we have found at large $D$, namely, $r_0/\sqrt{D}$ due to the decreasing area effect, and $r_0/D$ from the strong localization of the gravitational potential, the latter will be the most important one since we will see that it controls much of the classical black hole physics.

\subsection{Absence of interactions}\label{subsec:nointer}

Outside the horizon, over length scales of order $r_0$\, (\ie\ $r_0/D^0$), the gravitational potential $(r_0/r)^{D-3}$ vanishes exponentially fast in $D$. When $D\to \infty$ the lines of force are infinitely diluted in the infinite number of directions available, so there is no gravitational force outside the horizon. The horizon itself becomes a surface of infinite curvature. The spacetime of the black hole is then a flat geometry with a sphere cut off at the radius $r=r_0$. The area of this sphere is at a much smaller length scale.

This implies that when $D\to \infty$, at distances on the scale of $r_0$ there is no force at all between two black holes.\footnote{We stress that this sentence must be taken in the sense of parametric dependence on $D$: the distance could be, say, $r_0/100$, as long as it remains fixed as $D$ increases beyond $100$.} Multiple black hole solutions are obtained by simply cutting off spheres at different places in flat spacetime. 

The absence of a gravitational force on these scales, or equivalently the flatness of the metric outside the horizon, is also reflected in the fact that, for the solution \eqref{schwd}, we have that
\beq\label{GM}
GM=\frac{(D-2)\Omega_{D-2}}{16\pi}r_0^{D-3}
\eeq
vanishes as $D\to\infty$ due to the factor $\Omega_{D-2}$.\footnote{As explained above, if we are focusing on physics at the scale $r_0$ when taking $D\to\infty$, we could set $r_0=1$ or equivalently consider the dimensionless quantity $GM/r_0^{D-3}$.} 
A caveat is now in order. The classical theory of \eqref{Rmunu} 
is in essence a purely geometric theory that only contains geometrical magnitudes, such as lengths and areas. Any other physical magnitudes, such as mass or angular momentum, which require the introduction of conversion constants such as $G$ (absent from \eqref{Rmunu}), have only secondary meaning. Thus, $G$ and $M$ do not have any independent meaning in vacuum gravity. Clearly, all scales in vacuum gravity must have an interpretation of purely geometric origin. $GM$ can be regarded as a measure of how the extrinsic curvature of surfaces of constant $r$  differs from the extrinsic curvature of the same surfaces when embedded in Minkowski space. At large $D$, this difference vanishes and hence $GM\to 0$. We may consider introducing a gravitational mass-length scale
\beq\label{lmass}
\ell_M\sim (GM)^{1/(D-3)}\sim \frac{r_0}{\sqrt{D}}\,,
\eeq
characterizing this effect. This is the same as \eqref{larea}. 

Similarly, when coupling gravity to other matter systems, all the dimensionful parameters of the latter can be converted into length parameters. By choosing how they scale with $D$ we focus on specific regimes of gravitational physics. The conceptual prevalence of geometric magnitudes must be borne in mind whenever we choose, for physical illustration, to frame our discussion in terms of non-geometric quantities.

A case in point is the Bekenstein-Hawking entropy. Its definition requires the introduction of a new length scale not present in the classical theory, namely the Planck length $L_\mathrm{Planck}$, such that
\beq
S_{BH}=\frac{A_H}{4L_\mathrm{Planck}^{D-2}}\,.
\label{SAL}
\eeq
In fact, in the quantum vacuum theory $L_\mathrm{Planck}$ is the only dimensionful parameter that enters, \ie\ $G$ and $\hbar$ enter only through $G\hbar=L_\mathrm{Planck}^{D-2}$. Quantum effects on black holes are governed by the dimensionless ratio $r_0/L_\mathrm{Planck}$. We may keep it fixed as $D$ grows, or increasing with $D$ at a certain rate. Each of these choices specifies how large in Planck units are the black holes we are considering, \ie\ which quantum effects, if any, we want to focus on at large $D$. For instance, we may scale $r_0/L_\mathrm{Planck}\sim\sqrt{D}$  so as to keep the entropy 
\beq
S_{BH}\sim \left( \frac{r_0}{\sqrt{D}\,L_\mathrm{Planck}}\right)^{D}
\label{Sr0L}
\eeq
finite, but other choices may also appear natural. We will return to this point in sec.~\ref{sec:quantum}.

Note that if we consider ratios between entropies of black holes we eliminate the need to specify the scaling of $r_0/L_\mathrm{Planck}$. Then only the ratios of {\it classical} horizon areas are relevant (although it may still be convenient to talk in terms of entropies). This can be useful when comparing the initial and final states in a process. 
We shall do this now, in order to discuss further support for the picture of black holes as non-interacting objects. The Bekenstein-Hawking entropy of the Schwarzschild black hole behaves like
\beq
S(M)\sim M^{\frac{D-2}{D-3}}\xrightarrow{D\to\infty} M\,.
\eeq
In contrast to the situation at finite $D$, where $S\propto M^\alpha$ with $\alpha>1$, the fact that $S\propto M$ means that there is no entropic gain in merging two black holes. Nor is there any entropic penalty in splitting a black hole in two (recall that the horizon becomes singular in the limit). That is, 
\beq\label{Sratio}
\frac{S_{final}-S_{initial}}{S_{initial}}=\frac{A_{H,final}-A_{H,initial}}{A_{H,initial}}\to 0\,,
\eeq
both in the fragmentation or the merger of black holes.
This is a reflection of the absence of interactions noted above. 

Consider now a black $p$-brane,
\beq\label{bbrane}
ds^2=-f_p dt^2 +\sum_{i=1}^p dz_i^2 +f^{-1}_p
dr^2 +r^2 d\Omega_{D-p-2}\,,
\eeq
\beq
f_p=1-\lp\frac{r_0}{r}\rp^{D-p-3}\,.
\eeq
This brane is characterized by an energy density $\varepsilon$ and a pressure $P$ along its worldvolume such that
\beq\label{Pep}
P= -\frac{\varepsilon}{D-p-2}\,.
\eeq
When $D\to \infty$ this pressure vanishes: the brane has a dust equation of state (again, the precise statement is that $P/\varepsilon\to -1/D\to 0$).

The instability of the dust brane to fragmentation in this limit is easy to establish. At any finite $D$, neutral black branes suffer from a Gregory-Laflamme instability to growing inhomogeneities along the worldvolume, and are expected to eventually break up \cite{Lehner:2010pn}. The threshold mode at the onset of this instability has been studied at large $D$ in \cite{Kol:2004pn,Asnin:2007rw}, with the result that perturbations with wavelength longer than 
\beq\label{GLD}
\lambda_c=\frac{2\pi r_0}{\sqrt{D}}\lp 1+O(D^{-1})\rp
\eeq 
are unstable. Thus, when $D\to\infty$ perturbations of arbitrarily short wavelength drive the break up of the brane. Note that the effect occurs at the scale \eqref{larea}. In section~\ref{sec:GL} we will revisit this instability in much more detail, and discuss the appearance of the area-length scale \eqref{larea} in \eqref{GLD}.

For black $p$-branes we can regard $p$ as a parameter that can scale with $D$ in different manners. The case of $p\sim O(D^0)$ has been discussed above. Another possibility is that $p\sim D$ in such a way that $n=D-p-3$ remains finite, \ie\ we keep fixed the codimension of the brane instead of its worldvolume dimension. The gravitational field of these branes remains finite outside the horizon, and so does, too, their energy density and pressure. It seems appropriate to regard these black branes as belonging in a different sector of the theory than the infinitely localized ones we are considering so far, as they cannot be produced through processes involving a finite number of the latter. Their dynamics appears to be dominated by the degrees of freedom of a Kaluza-Klein reduction down to $n+3$ dimensions. Except for a brief mention in the next subsection, we will not consider these branes in the remainder.

\subsection{Other black holes: rotation and other topologies}\label{subsec:rotn}

When rotation is present in any $D\geq 4$ there are large classes of black hole solutions with many possible horizon topologies. Although the entire space of solutions has not been fully mapped, we can argue that the limit $D\to\infty$ results in configurations with  singular horizons and flat space outside them. Equilibrium horizons are equipotential surfaces, and at large $D$ the gravitational potential falls off steeply in the transverse direction away from them. If we take the limit $D\to\infty$ in such a way that the length scales that characterize the horizon remain finite, then the potential will drop infinitely fast outside the horizon. A possible concern is that in $D\geq 4$ there are horizons that are much more elongated in some directions than in others, and in these cases the gravitational potential close to the horizon falls off more slowly than at asymptotically large distances. Nevertheless, if the limit $D\to\infty$ is taken keeping the horizon length scales finite, the region where this slower fall off occurs shrinks and disappears in the limit. We can confirm all these features by studying explicit known solutions.

\paragraph{Myers-Perry black holes.}
Each independent rotation introduces a new length parameter $a$. Even if this parameter is usually regarded as the ratio of angular momentum to mass, as we discussed earlier the length $a$ is a more basic, purely geometrical quantity. We take these parameters $a$ to scale in such a manner as to preserve finiteness of the metric. 

Let us take the limit $D\to\infty$ for Myers-Perry black holes with several non-zero rotation parameters. We find no important distinction between the even- or odd-dimensional cases, so for definiteness we take odd $D$. The solution is \cite{Myers:1986un}
\beq 
ds^2=-dt^2
+(r^2+a_i^2)(d\mu_i^2+\mu_i^2 d\varphi_i^2) +\frac{r^2 r_0^{D-3}}{\Pi F}
(dt+a_i\mu_i^2 d\varphi_i)^2+\frac{\Pi F}{\Pi-r^2 r_0^{D-3}}dr^2\,, \label{even}
\eeq
where summation over $i=1,\dots,\frac{D-1}{2}$ is assumed, direction cosines satisfy $\mu_i^2=1$, and
\beq
F=1-\frac{a_i^2\mu_i^2}{r^2+a_i^2}\,,\qquad
\Pi=\prod_{i=1}^{(D-1)/2}(r^2+a_i^2)\,. \label{fpi}
\eeq
When $r_0=0$ the metric 
\beq\label{flata}
ds^2=-dt^2
+(r^2+a_i^2)(d\mu_i^2+\mu_i^2 d\varphi_i^2) +F dr^2
\eeq
describes flat space in ellipsoidal coordinates. When $r_0>0$, the horizon is at $r=r_H$ such that $r_H^2 r_0^{D-3}=\Pi(r_H)$. Let $k$ be the number of non-zero rotation parameters $a_i$, and for simplicity of illustration set them all equal to $a$. Then we have
\beq\label{rPI}
\frac{r^2 r_0^{D-3}}{\Pi}=\lp \frac{r_0}{r}\rp^{D-3}\lp 1+\frac{a^2}{r^2}\rp^{-k}\,,
\eeq
so $r_H$ is found by solving
\beq\label{rHa}
\frac{r_0}{r_H}=\lp 1+\frac{a^2}{r_H^2}\rp^\frac{k}{D-3}\,.
\eeq
If we keep $k$ fixed when $D\to\infty$ then $r_H\to r_0$, and $r^2 r_0^{D-3}/\Pi$ clearly vanishes for $r>r_0$: the metric becomes flat spacetime everywhere outside the horizon.

If instead the number $k$ of non-zero spins grows with $D$ such that $m=\frac{D-1}{2}-k$ remains fixed, then in the limit $D\to\infty$ eq.~\eqref{rHa} becomes
\beq
\frac{r_0}{r_H}=\lp 1+\frac{a^2}{r_H^2}\rp^{1/2}\; \Rightarrow\; r_H=\sqrt{r_0^2-a^2}\,.
\eeq
So $r_H\nrightarrow r_0$, but still
\beq
\frac{r^2r_0^{D-3}}{\Pi}\to \lp \frac{r_H^2+a^2}{r^2+a^2}\rp ^{D/2}
\eeq
vanishes, and thus leaves flat spacetime, at any $r >r_H$ when $D\to\infty$. This is the case even when $r_0\simeq a$, in which case $r_H$ is very small, the horizon can be highly pancaked (at least if some spins vanish), and $r_H\ll a$, which can be regarded as an ultraspinning limit~\cite{Emparan:2003sy}. The main point to note is that, in spite of the fact that at finite $D$ there is a region very close to the black hole where the potential falls off like $r^{-2(m-1)}$ (instead of $r^{3-D}$), the radial extent of this region vanishes in the limit $D\to\infty$ when we keep $r_0$ and $a$ fixed.

Summarizing, in the limit $D\to\infty$ the gravitational field vanishes completely outside the horizon, and the effect of $a\neq 0$ is that the surface in \eqref{flata} that is cut off at $r=r_0$ is not a sphere but an ellipsoid. 

The product $GM$ of the solution vanishes in the limit for the same reason as in the static case. The angular momentum in gravitational units, $GJ$, also vanishes. It is interesting that the ratio
\beq
\frac{J}{M}=\frac{2a}{D-2}\,,
\eeq
which is not suppressed by factors of $\Omega_{D-2}$, approaches zero. On the other hand the angular velocity on the horizon
\beq
\Omega_H=\frac{a}{r_H^2+a^2}
\eeq
remains finite when $D\to\infty$, even though no rotational dragging is felt anywhere outside $r=r_H$. The surface gravity in the singly-rotating case,
\beq
\kappa=\frac12\lp\frac{2r_H}{r_H^2+a^2}+\frac{D-5}{r_H}\rp
\eeq
diverges like it did in the static case, but its minimum at
\beq
\frac{a}{r_H}=\sqrt{\frac{D-3}{D-5}}
\eeq
remains finite as $D\to\infty$. This suggests that the value $a/r_H=1$ separates the regimes in which rotating black holes at large $D$ are stable or unstable to axisymmetric perturbations of the type discussed in \cite{Emparan:2003sy}. 


\paragraph{Blackfolds.}
Other types of rotating black holes are known to exist in $D\geq 5$. Exact analytical solutions have only been found in five dimensions, but approximate solutions are known for black rings and many other black holes in arbitrary $D\geq 5$ \cite{Emparan:2009at,Emparan:2009vd}. These are built by smoothly bending black branes, and are referred to as blackfolds. We can infer their large $D$ limits from the properties of black $p$-branes.

It is easy to see that a blackfold constructed out of a black $p$-brane, with $p$ fixed as $D\to\infty$, becomes a `dust-$p$-fold' as $D\to\infty$ when we keep finite the appropriate length parameters that determine its geometry (\eg\ horizon thickness and worldvolume curvature radius). For instance, for a black ring with horizon $S^1\times S^{D-3}$ we keep the radius of the $S^{D-3}$ and of the $S^1$ fixed as $D\to\infty$, and find a `ring of dust' with vanishing rotation velocity and a flat metric outside the singular horizon.

We may also have black $p$-folds for which $p$ diverges but $n=D-p-3$ remains fixed as $D\to\infty$. We are interested in the cases for which the gravitational potential beyond some distance $r_c$ remains bounded above and behaves, at finite $D$, like $(r_c/r)^{D-3}$. Then when $D\to\infty$ it will vanish for $r>r_c$. While it might seem possible that the radius $r_c$ lies at a finite distance outside the horizon, the evidence we have found from the Myers-Perry solutions, which include blackfolds in this class, confirms our expectation that the spacetime becomes flat outside the horizon. For these objects, $GM$ must vanish. A simple example besides the Myers-Perry solutions are homogeneous black $p$-folds with volume  $V_{D-n-3}$, for which \cite{Emparan:2009vd}
\beq
16\pi G M =(D-2)V_{D-n-3}\Omega_{n+1}r_0^n\,.
\eeq
We are keeping $n$ and $r_0$ fixed. Now, in the case of \eg\ a spherical blackfold, the volume $R^{D-n-3}\Omega_{D-n-3}$, and with it $GM$, vanishes due to $\Omega_{D-n-3}\to 0$. So blackfolds of this type appear to conform to our general picture. There can also be blackfolds for which $V_{D-n-3}$ does not vanish when $D\to\infty$, for example toroidal blackfolds with $V_{D-n-3}=L^{D-n-3}$ with $L$ fixed. Their mass-length \eqref{lmass}
\beq
\ell_M\sim L^{\frac{D-n-3}{D-3}}r_0^{\frac{n}{D-3}}(D-2)^{\frac1{D-3}}\to L
\eeq
remains finite. It is doubtful that these objects can be sensibly defined in the limit $D\to\infty$: at large distances their metric coefficients cannot be finite. If a suitable limit existed, they would form a different, infinitely more massive sector of the theory, similar to the black $p$-branes discussed at the end of sec.~\ref{subsec:nointer}. In this article we will not consider them any further.


\subsection{Charge}

The gravitational effect of the electric field in a Reissner-Nordstrom black hole is characterized by a charge-radius $r_Q$, related to the charge $Q$ by \cite{Myers:1986un}
\beq\label{Qcharge}
Q=\sqrt{\frac{(D-2)(D-3)}{8\pi G}}\,r_Q^{D-3}\,.
\eeq
This radius appears in the metric in the form $(r_Q/r)^{2(D-3)}$.
If, as may be natural in this context, we assume that we keep $r_Q$ fixed, then the gravitational effect of the charge vanishes at distances $r>r_Q$. Note that $Q$ is measured in units of the gauge coupling (the electron charge), so whether $Q$ remains finite in the limit depends on how this coupling scales with $D$.

Similarly, the effect of $p$-form charges (with $p$ not scaling with $D$), which can be carried by black $p$-branes or by black holes with non-spherical topologies \cite{Emparan:2004wy,Caldarelli:2010xz,Emparan:2011hg}, vanishes in the directions transverse to the brane if we keep their charge-radius fixed. However, a $p$-form charge forbids the break up of the brane in the $p$ directions parallel to the worldvolume. In these cases the equation of state of the brane does not become dust-like. For instance, for dilatonic $p$-branes in the limit $D\to\infty$ one finds
\beq
P=-\frac{N \sinh^2\alpha}{1+N \sinh^2\alpha}\,\varepsilon\,,
\eeq
where $\alpha$ is the charge-boost parameter (which remains fixed as $D\to\infty$ if the charge radius is fixed) and $N$ determines the dilaton coupling (see \eg\ \cite{Emparan:2011hg}). The field nevertheless vanishes outside the horizon.

\subsection{Large $D$ in Anti-deSitter}

The cosmological constant $\Lambda$ in Anti-deSitter spacetime introduces a curvature length scale.\footnote{The first part of this discussion applies also to de Sitter spacetime and to Schwarzschild-de Sitter black holes away from the Nariai limit.} In the limit $D\to\infty$, instead of fixing $\Lambda$ it is sensible to keep 
\beq
L=\sqrt{\frac{(D-1)(D-2)}{2|\Lambda|}}\sim \frac{D}{\sqrt{2|\Lambda|}}\,, 
\eeq
since this is the length that appears in the metric. If we kept $\Lambda$ fixed, we would have $L\to\infty$ and the effect of the cosmological constant would disappear from the geometry.

Again, we focus on the simplest black hole solution, which has the same form as \eqref{schwd} but with \cite{Witten:1998zw}
\beq
f(r)=1-\lp\frac{r_0}{r}\rp^{D-3}+\frac{r^2}{L^2}\,.
\eeq
The event horizon is not at $r=r_0$ but rather at $r_H<r_0$, the real positive radius where $f(r_H)=0$. At large $D$,
\beq\label{rHAdS}
r_H=r_0\lp 1-\frac{1}{D}\ln \lp 1+\frac{r_0^2}{L^2}\rp+ O\lp D^{-2}\rp \rp \,.
\eeq
Thus, $r_H\to r_0$ when $D\to\infty$. Like in flat space, in this limit the gravitational effect of the black hole outside this horizon vanishes. Black holes are again non-interacting particles, now in Anti-deSitter spacetime. However, although their area vanishes for fixed $r_0$ due to $\Omega_{D-2}\to 0$, these black holes are not always `small AdS black holes', since $r_H$ can still be either smaller or larger than $L$. Their temperature is \footnote{This assumes that $r_0/L_\mathrm{Planck}\sim D^0$. We may avoid specifying this by referring to the surface gravity instead of $T$.}
\beq
T=\frac{(D-1)r_H^2+(D-3)L^2}{4\pi L^2 r_H}
\eeq
so
\beq\label{TAdS}
T\to\frac{D}{4\pi r_0}\lp1+\frac{r_0^2}{L^2}\rp + O\lp D^{0}\rp\,.
\eeq
Although this diverges when $D\to \infty$ (like in flat spacetime), as a function of $r_0$ this temperature reaches a minimum at $r_0=L$. We may then regard black holes as small or large according to whether $r_0<L$ or $r_0>L$. We see in \eqref{rHAdS}, \eqref{TAdS} that the presence of $L$ modifies the large-$D$ short scales, in particular the surface gravity length is
\beq
\ell_\kappa \sim \frac{r_0 L^2}{D(r_0^2+L^2)}\,,
\eeq 
and its dependence on $r_0$ is rather different in the two regimes $r_0\lessgtr L$.

The shrinking effect of the Euclidean time circle and of the area of the spheres $S^{D-2}$  as $D\to \infty$ may be, if convenient, factored out by, \eg\ dividing $T$ by $D$, and dividing extensive thermodynamic quantities by $\Omega_{D-2}$. For instance, the Euclidean action of the black hole spacetime vanishes, both because of the factor $\Omega_{D-2}$ and because of the shrinking Euclidean time circle. In this case there is no trace of the Hawking-Page transition at $D\to\infty$. However, if we introduce a rescaled free energy
\beq
\hat F=\frac{16\pi G}{\Omega_{D-2}L^{D-3}}F= \lp\frac{r_H}{L}\rp^{D-3}
 \lp 1-\frac{r_H^2}{L^2}\rp\,,
\eeq
then the Hawking-Page transition at which $\hat F$ changes sign, is apparent at $r_H=L$, although extremely abrupt: at $D\to\infty$ this free energy changes from $-\infty$ for $r_H>L$, to $0^+$ for $r_H<L$. Another way to avoid these divergences is to consider ratios between extensive quantities.

The limit of very large AdS black holes yields planar black branes in Poincar\'e AdS. These black branes have a conformal equation of state 
\beq \label{Pepc}
P=\frac{\varepsilon}{D-1}\,,
\eeq 
so in the limit $D\to \infty$ they behave again like `dust' branes, which are marginally stable to longitudinal fluctuations, with empty AdS spacetime outside them. In section~\ref{subsec:fingl} we briefly revisit these branes.

\bigskip

In summary, all our knowledge about higher-dimensional black holes points to the conclusion that even when the effects of rotation, different horizon topologies, charge, or cosmological constant, are accounted for, then, on length scales for which the metric outside the horizon is finite, the limit $D\to\infty$ consists of configurations of objects with flat (or AdS) spacetime outside their singular horizons. We return now to extract more information from the basic solution \eqref{schwd}.

\section{Sphere of influence}\label{sec:sphinf}

While in the limit $D\to\infty$ the gravitational field vanishes at all $r>r_0$, on a scale $r_0/D$ there is a small region around the horizon where the black hole exerts its gravitational influence. More precisely, outside the horizon 
the gravitational potential is still appreciable, \ie
\beq
\lp\frac{r_0}{r}\rp^{D-3}=O(D^0)\,,
\eeq
within the region
\beq\label{infsph}
r-r_0\lesssim \frac{r_0}{D}+O(D^{-2})\,,
\eeq
while it vanishes exponentially fast in $D$ outside this range.

For some observables this `sphere of influence', characterizing the degree to which a phenomenon is localized close to the horizon, may extend out to a further radius. Typically this is due to radial derivatives of $f(r)$, each of which brings in a factor $\sim D$.
In general, the relevant result is that, for $r>r_0$,
\beq\label{gensph}
D^b\lp\frac{r_0}{r}\rp^{D}=O(D^0)\quad \Leftrightarrow\quad  r-r_0\lesssim \frac{r_0}{D}\lp a+ b\ln D\rp+O(D^{-2})\,,
\eeq
where $a,\, b$ are $D$-independent numbers. For instance, the effects of the gravitational force $\partial_r f$ have a range of this form with $b=1$. We see this in the acceleration of an observer at constant $r$ in \eqref{schwd},
\beq
a=\frac{D-3}{2r_0\sqrt{f(r)}}\lp \frac{r_0}{r}\rp^{D-2}\xrightarrow{D\to\infty} 
\frac{D}{2r_0}\lp \frac{r_0}{r}\rp^{D}\,,
\eeq
which has a range 
\beq
r-r_0\lesssim \frac{r_0}{D}\ln D+O(D^{-2})\,.
\eeq
An example with $b=2$ is the Kretschmann scalar \eqref{krets}, which is of order one in the region
\beq
r-r_0\lesssim \frac{2r_0}{D} \ln D+O(D^{-2})\,.
\eeq

The scattering of light rays thrown towards the black hole  \eqref{schwd} provides another illustration. These rays will be absorbed if the impact parameter $b$ is smaller than the critical value \cite{Emparan:2000rs}
\beq
b_c=\left(\frac{D-1}{2}\right)^{1/(D-3)} \sqrt{\frac{D-1}{D-3}}\,r_0\,.
\eeq
As $D\to\infty$ this gives $b_c\to r_0$. Light rays that pass with $b>r_0$ suffer no deflection: it is infinitely difficult to catch a line of force that pulls towards the black hole. The photon sphere of influence is at a scale $r_0/D$ around the horizon
\beq
b_c=r_0\left( 1+\frac{1+\ln(D/2)}{D}+O(D^{-2})\right)\,.
\eeq
The photon absorption cross section
\beq\label{bcrit}
\sigma=\frac{\Omega_{D-3}}{D-2}b_c^{D-2}
\eeq
is at the scale $(\ell_A)^D$, as already discussed.
For the rotating black hole \eqref{even}, the critical impact parameter for light rays in the plane $\mu_k=1$ asymptotes to $b_c\to \sqrt{r_0^2+a_k^2}$, which is precisely the radius of the circle at $r=r_0$ in this plane in the limiting flat geometry \eqref{flata}.

\subsection{Quasinormal modes}

Quasinormal modes characterize the black hole's own dynamics. It is natural to expect that they are localized close to the sphere of influence and controlled by the scale $r_0/D$. This is actually the case. At large $D$ the quasinormal frequencies can be estimated or computed (see \cite{Konoplya:2003ii,Berti:2003si,Berti:2009kk} and section~\ref{sec:scalar} below) with the result that for either scalar, vector or tensor perturbatons,
\beq \label{qnfreq}
\mathrm{Re}\,\omega_\mathrm{QN}\sim \frac{D}{r_0}\,,
\eeq 
at least when the angular momentum and overtone numbers are $\ll D^2$.
In the WKB approximation these quasinormal modes can be regarded as localized close to the circular photon orbit at
\beqa\label{photsph}
r_{ph}&=&\left(\frac{D-1}{2}\right)^{1/(D-3)}\,r_0\nonumber\\
&=& r_0\lp 1+\frac{\ln(D/2)}{D}\rp+O(D^{-2})\,.
\eeqa

The imaginary part of $\omega_\mathrm{QN}$ is suppressed at large $D$ relative to the real part, which may be related to the apparent ease with which black holes can fragment in this limit. 

\subsection{Across the horizon, a small interior}

In the limit $D\to\infty$ a particle that falls in the black hole encounters a singularity at $r=r_0$ (\ie\ a divergence for measurements on the scale of $r_0$) where its trajectory comes to an end. However, on the scale $r_0/D$ the horizon is a non-singular place and we can ask how long it takes for the particle to reach the singularity at $r=0$ --- \ie\ what scale controls the internal size of the black hole. 

Taking for simplicity a particle that starts at rest at infinity in a radial trajectory, the proper time elapsed between the moment it crosses the radius $r=R$ until it reaches the singularity is
\beq\label{tsing}
t_{sing}=\int_0^{R}dr\lp\frac{r}{r_0}\rp^\frac{D-3}{2}=\frac{2r_0}{D-1}\lp\frac{R}{r_0}\rp^\frac{D-1}{2}\,.
\eeq
The two factors in this result each have a different origin.
In the large $D$ limit the particle takes a time that diverges exponentially with $D$ to go from any finite distance outside the sphere of influence until it reaches this region, where $(R/r_0)^D\sim 1$; this is just a reflection of the very small gravitational force outside this sphere. From the moment when the particle enters this region until it reaches the singularity, the time \eqref{tsing} that passes is
\beq\label{tsing2}
t_{sing}\simeq \frac{2r_0}{D} +O(D^{-2})\,,
\eeq
\ie\ a very short time, determined once again by the scale $r_0/D$. Most of this time is spent in the sphere of influence, which now includes an inner region $r_0-r\lesssim r_0/D$. After that, the singularity is reached exponentially fast in $D$. Eq.~\eqref{tsing2} gives a sense in which we can regard the black hole interior as small relative to the scale $r_0$.

\section{Classical radiation from black hole interactions}\label{sec:radn}

The absence of a gravitational field inbetween two or more black holes implies that, when they move in the presence of one another, no gravitational radiation is emitted. Below we verify that this is consistent with the large $D$ limit of several previously studied processes, but there is an important effect that remains in phenomena that probe very short lengths~$\lesssim~r_0/D$.

\subsection{Black hole collisions}\label{subsubsec:colls}

For simplicity we study collisions among black holes of equal size.
It is clear that when $D\to \infty$ two black holes thrown towards one another will not be deflected from their straight paths unless the impact parameter is $\leq r_0$. An often studied case is the ultrarelativistic limit, \ie the collision of two Aichelburg-Sexl shockwave solutions. It is immediate to see that at $D\to\infty$ these solutions become trivially flat outside a singularity at the center of the shockwave plane --- away from this point, the shock itself disappears. So, in this limit, for any collision course that is not exactly head-on, we conclude, unshockingly, that the particles simply fly by each other, with no emission of radiation.

Put now the black holes in a head-on collision course. Collision is unavoidable, but will there be any radiation emission? If the black holes start towards each other with infinitesimal velocities, then our argument about the entropy in a merger in sec.~\ref{subsec:nointer} tells us that no radiation will be produced when they merge. More generally, if they move towards each other with initial velocities $\pm v$, then the area theorem imposes an upper bound on the ratio of the radiated energy to the initial energy \cite{Hawking:1971tu} which, when $D\to\infty$, is\footnote{Like in \eqref{Sratio}, we consider ratios of energies so that this is a finite quantity at $D\to\infty$ that can be defined in terms of purely geometric quantities.}
\beq\label{epsarea}
\epsilon(v)=\frac{E_{rad}}{E_{in}}\leq 1-\sqrt{1-v^2}\,.
\eeq
As $v\to 0$ we recover $\epsilon\to 0$, \ie\ no radiation, but for $v>0$ there is the possibility that radiation is produced. This indicates that kinetic energy, but not any rest mass of the black holes, can be converted into radiation.

We may expect that the actual $\epsilon$ is maximized in ultrarelativistic collisions and in fact when $v\to 1$ the condition \eqref{epsarea} imposes no constraint. For shockwave collisions a more stringent bound on $\epsilon$ can be obtained from the area of the apparent horizon at the moment that the shockwaves meet \cite{Eardley:2002re,Yoshino:2002br}. When $D\to\infty$ this bound becomes
\beq
\epsilon_\mathrm{shock}\leq \frac12\,.
\eeq
According to refs.~\cite{Coelho:2012sya,Coelho:2012sy}, this bound is actually saturated, and thus gravitational radiation is indeed produced at $D\to\infty$. This might appear to run against the picture of $D\to\infty$ black holes as non-interacting dust particles, but there is a plausible interpretation for this result. A head-on collision is an extremely fine-tuned process (even more so at $D\to\infty$ where the cross section scale \eqref{larea} shrinks to zero) which can excite dynamics at arbitrarily short distances  involving modes of arbitrarily high frequency, such as the quasinormal modes with $\omega\sim D/r_0$, which are localized in the region \eqref{photsph} very near the black holes. At large $D$ these modes can get excited if (and only if) very short scales $\sim r_0/D$ are probed. 

More precisely, the emission of radiation comes with a factor $\sim \omega^{D}$ from the frequency-volume available to radiation. If the characteristic length of modes that are radiated is $\ell$, then the process will be governed by a factor $(\omega\ell)^D$. 
Emission will be completely shut off when the frequencies that can be excited in a process are $\omega<\ell^{-1}$, but it may remain appreciable as $D\to\infty$ when $\omega\sim\ell$, even if $\ell$ vanishes in this limit. In the following we find evidence that this effect is indeed present, and moreover that the relevant scale is $\ell\sim r_0/D$, \ie\ that of typical quasinormal modes.\footnote{\label{foot:gravitons}The fact that the number of graviton polarizations grows like $D^2$ does not seem to play more than a subleading role, negligible for the previous argument.} 

Ref.~\cite{Cardoso:2002pa} estimated the amount of radiation produced in black hole collisions in the `instantaneous collision approximation' (which may in fact become more accurate as $D$ grows). At moderate (not ultrarelativistic) initial velocities, and ignoring $D$-independent numbers, the leading large $D$ emission per solid angle is\footnote{\label{fn}The dependence on dimensionful parameters in this equation can be easily worked out from generic considerations. But note that we are also accounting for the dominant $D$-dependent dimensionless factors.}
\beq
\frac{dE_{rad}}{d\Omega}\sim G M^2\omega_m^{D-3}\,,
\eeq
where $\omega_m$ is a (physical) cutoff in the frequencies that are radiated. Integrating over all angular directions, the ratio of the total radiated energy to the black holes' mass is
\beq
\epsilon\simeq\frac{E_{rad}}{2M}\sim \frac{\Omega_{D-2}}{M}\frac{dE_{rad}}{d\Omega} \sim
\lp \frac{\omega_m r_0}{D}\rp^D
\eeq
(we have used \eqref{lmass}). If the frequency cutoff were $\omega_m\sim r_0^{-1}$ then no radiation would be emitted when $D\to\infty$. But precisely at the frequencies $\omega_m\sim D/r_0$ there is a big enhancement in the emission of radiation. 

Our next example provides another instance in which there is a possibility of producing radiation by exciting frequencies $\sim D/r_0$. It is worth stressing that neither of these calculations involve a quasinormal mode study, nor indeed any perturbation analysis of the black hole spacetime \eqref{schwd}. We find remarkable the consistent appearance of this scale.

\subsection{Orbiting black holes}\label{subsubsec:orbs}

The radiating power from two black holes, each with Schwarzschild radius $r_0$ and mass $M$, orbiting around each other at a distance $l$ and with orbital frequency $\omega$, tends at large $D$ to~\footnote{See footnote \ref{fn}.}
\beq
\frac{dE_{rad}}{dt}\sim \Omega_{D-2} GM^2 l^4 \omega^{D+2}
\eeq
(again up to $D$-independent numerical factors) \cite{Cardoso:2002pa}. The fraction of the system's energy emitted per orbit is
\beq\label{radorb}
\epsilon\simeq \frac{\pi}{M\omega}\frac{dE_{rad}}{dt}\sim \left(\frac{\sqrt{D}l}{r_0}\right)^4 \left(\frac{\omega r_0}{D}\right)^D\,.
\eeq
If the black holes follow Keplerian orbits with $l>r_0$ then 
\beq
\omega\sim \sqrt\frac{GM}{l^{D-1}}\sim \frac1{l}\lp\frac{r_0}{\sqrt{D}l}\rp^{D/2}
\eeq 
vanishes as $D$ grows, since their mutual attraction becomes increasingly weaker (this also justifies the simplifications made in deriving \eqref{radorb}, although note that the orbits are unstable). This does fit the  picture of non-interacting, non-radiating particles. 

However, let us push the argument further and consider that the black holes follow circular trajectories under an external force (whose gravitational effect we neglect; we proceed even it is unclear how sensible this is), so that $\omega$ is independent of the black hole masses and their separation. The dominant factor in \eqref{radorb} at large $D$ is $(\omega r_0/D)^D$. So when the orbital frequency is $\omega \sim  D/r_0$, \ie\ when the motion is so fast that it is capable of exciting the quasinormal modes of the black hole, radiation is possible as $D\to\infty$. In this regime the approximations made in deriving \eqref{radorb} certainly cease to be justified, but still we can observe how the excitation of the very high-frequency modes in the thin region near the horizon may result in radiation that survives the limit $D\to\infty$.

\medskip

The themes of this section will reappear in sec.~\ref{sec:scalar}, where we confirm this picture of the interaction of the black hole with classical waves.

\section{Quantum effects}
\label{sec:quantum}

We do not study quantum gravitational effects in any detail but merely make some elementary remarks.

As we discussed after eq.~\eqref{SAL}, the relevance of quantum effects at large $D$ is determined by the $D$-dependence of the dimensionless ratio $r_0/L_\mathrm{Planck}=r_0/(G\hbar)^{1/(D-2)}$. We could set $L_\mathrm{Planck}=1$ without loss of generality and then discuss how $r_0$ scales with $D$ (or viceversa), but we prefer to keep both quantities explicit.

For Hawking radiation, the main parameter is the size of the length scale $\ell_\kappa$ \eqref{elk} ---or the length of the Euclidean time circle--- measured in units of $L_\mathrm{Planck}$.
In terms of dimensionless quantities, we write the Hawking temperature $T_H$ as
\beq\label{THEP}
\frac{T_H}{E_\mathrm{Planck}}=L_\mathrm{Planck}\frac{D-3}{4\pi r_0}\,,
\eeq
where $E_\mathrm{Planck}=\hbar/L_\mathrm{Planck}=L_\mathrm{Planck}^{D-3}/G$.
Thus the appropriate measure of the Hawking temperature is the value of $D L_\mathrm{Planck}/r_0$. An equivalent discussion can be made for the effect of higher-curvature corrections to the semiclassical gravitational effective theory in the close vicinity of the horizon, since these are also controlled by the scale  $\ell_\kappa$, see \eqref{lK}.

Eq.~\eqref{THEP} implies that when $r_0$ is, parametrically in $D$, at the same scale as $L_\mathrm{Planck}$, then the temperature becomes superPlanckian at sufficiently large $D$.
But we may also consider black holes with 
\beq\label{fixTH}
r_0\sim D L_\mathrm{Planck}\,.
\eeq
These have a big radius in Planck-length units, which makes their entropy~\eqref{Sr0L}
very large. Their temperature is fixed as we send $D\to\infty$ and thus it can remain safely below the Planck energy.\footnote{If $r_0/D\sim L_\mathrm{Planck}$ then the `sphere of influence' is of Planckian size parametrically in $D$. This suggests that it is consistent to think of it as a `stretched horizon', although note that $r_0/D$ can still be much larger than $L_\mathrm{Planck}$ by a $D$-independent large factor.}

The situation, however, is subtle. Hawking radiation at large $D$ was studied in ref.~\cite{Hod:2011zzb}. One might expect that the typical energy $\hbar\omega$ of Hawking quanta should be of the order of $ T_H$. However, the actual typical energies are much larger. The reason is the huge increase in the phase space available to high-frequency quanta at large $D$ (already encountered in sec.~\ref{sec:radn}), which shifts the radiation spectrum towards energies much larger than $T_H$. The number density distribution of quanta at temperature $T$ is
\beq\label{nomega}
n(\omega)d\omega=\Omega_{D-2}\frac{\omega^{D-2}d\omega}{e^{\hbar\omega/T}-1}\,,
\eeq
with the factor $\Omega_{D-2}$ coming from the angular integration in momentum-space. 
This density peaks at \cite{Hod:2011zzb}
\beq
\hbar\omega= DT\lp 1+O(e^{-D})\rp\gg T\,.
\eeq
Then, for a black hole the typical frequency of Hawking quanta is
\beq\label{peakom}
\omega_H\simeq D T_H/\hbar\simeq \frac{D^2}{4\pi r_0}\,.
\eeq
Equivalently, we can say that Hawking radiation probes distances extremely close to the horizon, at a scale $r_0/D^2$.
We see that even for the large black holes \eqref{fixTH} that have finite $T_H$,  Hawking radiation is emitted at superPlanckian energies 
\beq
\hbar\omega_H\sim D E_\mathrm{Planck}\gg E_\mathrm{Planck}\,.
\eeq 
In order to render the radiation subPlanckian and bring the effect under semiclassical control one should consider black holes of much larger size, $r_0\sim D^2 L_\mathrm{Planck}$, with $T_H\sim E_\mathrm{Planck}/D\ll E_\mathrm{Planck}$. 

Thus the black hole is a very \textit{large} quantum radiator, whose radius ($\sim r_0$) and typical classical vibrational wavelengths ($\sim r_0/D$) are much longer than the wavelengths it radiates quantum mechanically ($\sim r_0/D^2$). 
In this respect it is interesting that, even if phase space integrals receive suppressing factors  of $\Omega_{D-2}$ from angular integrals (from both momentum and position space), these are offset by the growth $\sim \omega^{D-2}$ of volume in radial (frequency) directions. The number density \eqref{nomega} at the peak frequency \eqref{peakom} behaves like
\beq
n(\omega_H)\sim \Omega_{D-2} (T_H D/\hbar)^{D-2} \sim D^{3D/2}r_0^{-D}\,.
\eeq
Then, the total power $P$ of Hawking flux radiated by an object of area $A_H\sim D^{-D/2}r_0^D$, still shows factorial growth, $P\sim D^D E_\mathrm{Planck}/r_0$. This enhancement is due to the fact that $T_H\sim D \hbar/r_0$.\footnote{See footnote~\ref{foot:gravitons}.}
For practical applications of the large $D$ limit, the ultrashort wavelength of Hawking radiation can be a bonus since it implies that the geometric optics approximation for graybody factors applies very accurately \cite{Hod:2011zzb}. 

It would be interesting to investigate whether the black hole information problem can be sensibly formulated in the $1/D$ expansion. Let us mention, among the potentially relevant factors, that observers performing experiments on a semiclassical black hole, $r_0/L_\mathrm{Planck} \geq O(D^0)$, using probes at the natural `outside scale' $r_0$, perceive the horizon as a singular surface which perfectly reflects all radiation of frequency $\sim 1/r_0$. This might seem to prevent that the information of such outside matter be lost across the horizon (although it might also be destroyed there), while Hawking pairs come out at a much higher frequency $\sim D^2/r_0$. The very short time \eqref{tsing} available for measurement for observers who cross the horizon might facilitate the consistency of a form of black hole complementarity, but the extremely fast evaporation rate compresses enormously \textit{all} the timescales involved. It is then unclear without a more careful investigation whether these are positive or negative features for making this a useful approach. 

\section{Large $D$ effective theory: scalar wave absorption}
\label{sec:scalar}

Up to this point we have been mostly drawing consequences from the limit $D\to\infty$ in processes that had already been calculated at finite $D$. We have gathered ample evidence that a consistent picture emerges. 
Now we apply the large $D$ expansion to problems that have not been fully solved previously in analytic form. The main idea that simplifies their study is the following.

\subsection{Near and far regions}\label{subsec:nearfar}

The gravitational field is strongly localized within a region close to the horizon characterized by the scale $r_0/D$. This is a length that at large $D$ is widely separate from the radius $r_0$, so we can define two distinct but overlapping regions in the geometry:
\beqa
&\mathrm{near~region:} & r-r_0\ll r_0\,, \nonumber\\
&\mathrm{far~region:} & r-r_0\gg \frac{r_0}{D}\,. 
\eeqa
Equivalently, introducing the variable
$ \sR\equiv(r/r_0)^D$
they can be characterized as
\beqa
&\mathrm{near~region:} & \ln \sR\ll D \,, \nonumber\\
&\mathrm{far~region:} & \ln \sR\gg 1 \,.
\eeqa
The far region, where the geometry is effectively flat, excludes the sphere of influence \eqref{infsph}. The latter is part of the near region,\footnote{Incidentally, the near region retains more features than a Rindler horizon.} which extends into an 
\beq
\mathrm{overlap~region:}\quad  \frac{r_0}{D}\ll r-r_0\ll r_0 \,,\quad \mathrm{\ie}\quad
1\ll\ln \sR\ll D\,,
\eeq 
very close to the horizon, where it smoothly connects to the far region. Then,  the study of {\it any} phenomenon that takes place in this geometry lends itself naturally to the method of matched asymptotic expansions (first used in this context in \cite{Asnin:2007rw}). If we manage to solve the field equations in the near region, with regularity conditions imposed on the horizon, then this solution will provide boundary conditions for the far field by requiring that the fields match where they overlap.

\subsection{Massless scalar wave equation}\label{sec:scalarwave}

We study massless scalar field propagation in the black hole background \eqref{schwd}. 
Throughout this section it will be slightly convenient to work with the parameter
\beq
n=D-3
\eeq
instead of $D$. At large $D$ they are of course equivalent.

%
We study the wave equation
\begin{equation}
\Box \Psi \,=\, 0 \label{Psieq}
\end{equation} 
%
in the background \eqref{schwd}. We set 
\beq
\Psi=e^{-i\omega t}\, \psi_{\omega l}(r)\, Y^{(l)}_{n+1}(\Omega)\,,
\eeq 
where $Y^{(l)}_{n+1}(\Omega)$ are spherical harmonics on $S^{n+1}$, and henceforth we drop the mode indices from $\psi$. The equation becomes
%
\begin{equation}
\frac{1}{r^{n+1}}\frac{d}{dr}\left( r^{n+1} f(r) \frac{d}{dr}\psi \right) +\frac{\omega^{2}}{f(r)}\psi -\frac{l(l+n)}{r^{2}}\psi \,=\,0\,. \label{psieq}
\end{equation} 
%
It is conventional to write this as an equation for  
\beq\label{phipsi}
\phi(r)=r^{(n+1)/2}\psi(r)
\eeq 
in the form
\begin{equation}
\frac{d^{2}\phi}{dr_{*}^{2}}+\left(\omega^{2}-V(r_*)\right)\phi\,=\,0, \label{psieq2}
\end{equation} 
where the tortoise coordinate defined by $dr_{*} =dr/f(r)$ is given in terms of a hypergeometric function,
\beq
r_{*}={}_{2}F_{1}\left( -\frac1n,1,\frac{n-1}{n};\lp\frac{r_0}{r}\rp^n\right)\,r \,,
\eeq
and
%
\begin{eqnarray}
V(r_{*})\,=\frac{f(r)}{4r^2}\left( (2l+n)^2-1+(n+1)^{2}\left(\frac{r_{0}}{r}\right)^{n} \right)
\label{poten}
\end{eqnarray}
%
is the Regge-Wheeler potential for these perturbations.

At large $n$ this potential scales like $n^2$. In order to capture physics of interest we introduce \footnote{Ref.~\cite{Kol:2011vg} also notes that $\hat l$ can be more relevant than $l$ when $D$ is regarded as a parameter.}
\beq\label{omeganln}
\hat\omega=\frac{\omega}{n}\,,\qquad
\hat l=\frac{l}{n}
\eeq
and consider $\hat\omega$ and $\hat l$ as $O(1)$ quantities. Now when $n$ is large the equation becomes
\beq\label{psieq3}
\frac{d^{2}\phi}{dr_{*}^{2}}+n^2\left(\hat\omega^{2}-\hat V(r_*)\right)\phi=0\,,
\eeq
with
\beq
\hat V(r_*)=\frac{f(r)}{4r^2}\left( (2\hat l+1)^2+\left(\frac{r_{0}}{r}\right)^{n} \right)\,.\label{hatpot}
\eeq
This potential vanishes on the horizon at $r_*\to-\infty$ and in the asymptotic region at $r_*\to+\infty$. It has a maximum at
\beq
r=r_\mathrm{max}=r_0\lp 2n\frac{\hat l(\hat l +1)}{(2\hat l+1)^2}+O(n^0)\rp^{1/n}\,.
\eeq
When $\hat l\gg 1$ this is
\beq
r_\mathrm{max}=r_0\lp 1+\frac{\ln(n/2)}{n}\rp+O(n^{-2})\,,
\eeq
which reproduces the radius of the circular photon orbit \eqref{photsph}.
The maximum of the potential, corresponding to critical scattering at the threshold of absorption, occurs for the frequency $\hat\omega=\omega_c$, with
\beq\label{omcrit}
\omega_{c}r_0= \hat l+\frac12+O(n^{-1})\,.
\eeq
This gives the real part of the quasinormal mode frequency \eqref{qnfreq} in the WKB approximation. The critical impact parameter $b_c=\hat l/\omega_c$ at $\hat l\gg 1$ reproduces the geometric optics result \eqref{bcrit} at large $n$. This critical frequency will play a central role in the problem. 

When $n\to\infty$ this potential becomes extremely simple: for $r_*>r_\mathrm{max}$ (where $r_*= r$) we recover the flat space potential with only a centrifugal barrier. For $r_*<r_\mathrm{max}$ the potential vanishes exponentially quickly in $n$. So the limiting form of the potential is
\beq\label{limhatpot}
\hat V(r_*)\to \frac{\omega_c^2 r_0^2}{r_*^2}\,\Theta(r_*-r_0)\,.
\eeq

We recognize here two of the main running themes of this article. First, waves outside the black hole propagate in a flat space potential. Its height $V(r_\mathrm{max})\sim n^2$ becomes infinite when $n\to\infty$, so in this limit excitations of frequency $\omega=O(n^0)$ are not absorbed at all: this radiation does not interact with the black hole. Second, waves of frequency $\omega \gtrsim n/r_0$ can penetrate the barrier and probe the dynamics near the horizon of the black hole.

Problems of field propagation are characterized by the field amplitudes at infinity and at the horizon,
\beq
\phi \sim \begin{cases}\, A^\mathrm{in}_l(\omega)\lp e^{-i\omega r_{*}} +R_l(\omega) e^{i\omega r_{*}}\rp &\quad r_{*}\rightarrow \infty\,, \\
\, A^\mathrm{in}_l(\omega) T_l(\omega) e^{-i\omega r_{*}}    &\quad r_{*} \rightarrow -\infty\,.\end{cases} \label{bc}
\eeq
When the amplitude of the incoming wave $A^\mathrm{in}_l(\omega)$ is nonzero, the physical information is contained in the reflection and transmission (\ie\ absorption) amplitudes, $R_l(\omega)$  and $T_l(\omega)$. We will solve this problem of scattering and absorption by the black hole, and in particular compute the absorption probability
\beq
\gamma_l(\omega)=|T_l(\omega)|^{2}\,.
\eeq
The condition $A^\mathrm{in}_l(\omega)=0$ defines a different kind of problem: the determination of the spectrum of quasinormal modes of the black hole, which appear as poles in $T_l(\omega)$ and $R_l(\omega)$. This is a subtle eigenvalue problem to solve in the near region, and we postpone its detailed study to future work.

\subsection{Integrating out the near region}

In order to simplify the notation, we now set
\beq
r_0=1,
\eeq
so that
\beq
f(r)=1-r^{-n}\,.
\eeq

The tortoise coordinate $r_*$ is not quite appropriate in the near region $-\infty <r_*<r_\mathrm{max}$: the non-trivial features of the potential \eqref{hatpot} in this region are erased in the limit \eqref{limhatpot} in which $r_*$ remains fixed as $n\to\infty$. Although we could instead keep $\hat r_* = n r_*$ fixed, we find more useful to work with the coordinate 
\beq
\sR=r^{n},\label{sR}
\eeq 
in terms of which the wave equation \eqref{psieq} becomes
%
\begin{equation}
\frac{d}{d\sR}\left( \sR(\sR-1)\frac{d}{d\sR}\psi \right) -\hat{l}(\hat{l}+1)\psi +\hat{\omega}^{2}\frac{\sR^{1+2/n}}{\sR-1}\psi \,=\,0\,. \label{psieqpre}
\end{equation} 
To leading order at large $n$ in the near region, where $\ln\sR\ll n$, this equation becomes
%
\begin{equation}
\frac{d}{d\sR}\left( \sR(\sR-1)\frac{d}{d\sR}\psi  \right) -
\lp\omega_c^2-\frac14\rp\psi 
+\hat{\omega}^{2}\frac{\sR}{\sR-1}\psi \,=\,0,
\end{equation} 
%
where instead of $\hat l$ we use $\omega_c$, defined in \eqref{omcrit}. The expansion breaks down when $\hat\omega$ is of order $n$ or higher. Then our calculations apply in the range $\omega\ll n^2$. 

The general solution to this equation is
%
\begin{eqnarray}
\psi(\sR)&=&A_{1}(\sR-1)^{-i\hat{\omega}}~{}_{2}F_{1}(q_{+},q_{-},q_{+}+q_{-};1-\sR) \notag \\
&&+A_{2}(\sR-1)^{i\hat{\omega}}~{}_{2}F_{1}(1-q_{+},1-q_{-},2-q_{+}-q_{-};1-\sR), \end{eqnarray} 
%
where
%
\begin{equation}
q_{\pm}\,=\,\frac{1}{2}-i\hat{\omega}\pm \sqrt{\omega_c^{2}-\hat{\omega}^{2}}\,. \label{qdef}
\end{equation} 
%
The regularity (ingoing) boundary condition at the horizon (\ref{bc}) takes the form
%
\begin{equation}
\psi(\sR)\Big|_{\sR=1}\,\propto \,(\sR-1)^{-i\hat{\omega}}\left(1+O(n^{-1}) \right)
\end{equation} 
%
and therefore requires that $A_{2}=0$. We use the arbitrariness in the overall amplitude to set $A_1=1/\sqrt{n}$ for later convenience.

The solution can now be written as
%
\begin{eqnarray}\label{nearsol}
\psi(\sR)&=&\frac{\Gamma(q_{+}+q_{-})}{\sqrt{n}}(\sR-1)^{-i\hat{\omega}}
\Bigg[ \sR^{-q_{-}}\frac{\Gamma(q_{+}-q_{-})}{\Gamma(q_{+})^{2}}
\,{}_{2}F_{1}(q_{-},q_{-},1-q_{+}+q_{-};1/\sR) \notag \\
&&\qquad\qquad\qquad\qquad+
\sR^{-q_{+}}\frac{\Gamma(q_{-}-q_{+})}{\Gamma(q_{-})^{2}}
\,{}_{2}F_{1}(q_{+},q_{+},1+q_{+}-q_{-};1/\sR)\Bigg].
\end{eqnarray} 
%
Expanding at large $\sR$ yields the solution in the overlap region  $1\ll \ln{\sR}\ll n$,
%
\begin{eqnarray}
\psi(\sR)=\frac{\Gamma (q_{+}+q_{-})}{\sqrt{n}}\sR^{-\frac{1+q_{+}+q_{-}}{2}}\biggl(
\frac{\Gamma(q_{+}-q_{-})}{\Gamma(q_{+})^{2}}\sR^{q_{+}}+
\frac{\Gamma(q_{-}-q_{+})}{\Gamma(q_{-})^{2}}\sR^{q_{-}}
\biggr)+O(\sR^{-1})\,. \label{nearexp}
\end{eqnarray} 
%
When $q_+=q_-$, \ie\ $\hat\omega=\omega_c$, this expansion is not valid, and instead one gets terms $\sR^{-1/2}\ln\sR$. We expect that this is due to the presence of quasinormal modes. Logarithmic terms also appear when $q_+-q_-\in\mathbb{N}$.

The flux of the scalar at the horizon, derived from the wave equation \eqref{psieq2}, is
%
\beqa
F_{\text{horizon}}\,&=&\,\frac{i}{2}\left( \phi^{*}\frac{d}{dr_{*}}\phi-\phi\frac{d}{dr_{*}}\phi^{*} \right)\Big|_{r_* \to-\infty}\notag\\ \notag\\
\,&=&\,\frac{i n}{2}\sR(\sR-1)\lp \psi^* \frac{d}{d\sR}\psi-\psi \frac{d}{d\sR}\psi^*\rp\Big|_{\sR=1}\,.
\eeqa
%
For our solution (\ref{nearsol}) we find
%
\begin{equation}
F_{\text{horizon}}\,=\,\hat\omega.
\end{equation} 
%

Eq.~\eqref{nearexp} is an important result in this analysis: by providing a boundary condition for the fields that propagate outside the `sphere of influence', it codifies the physics of the region where all the black hole dynamics is concentrated.

\subsection{Far region waves}

In the far region we set $f(r)\rightarrow 1+O(e^{-n})$: the wave propagates effectively in flat space, obeying the equation 
%
\begin{equation}
\frac{1}{r^{n+1}}\frac{d}{dr}\left( r^{n+1}\frac{d}{dr}\psi(r)\right)+n^2\left(\hat\omega^{2}-\frac{\hat l(\hat l+1)}{r^{2}}\right)\psi(r)\,=\,0\,.
\end{equation} 
%
This is solved in terms of Bessel functions,
%
\begin{equation}\label{farscsoln}
\psi(r)\,=\,C_{1}\frac{J_{n\omega_c}(n\hat\omega r)}{r^{n/2}}+C_{2}\frac{Y_{n\omega_c}(n\hat\omega r)}{r^{n/2}}.
\end{equation} 
%
From the behavior at large $r$ 
%
\begin{eqnarray}
\psi(r)\,\simeq\,\frac{1}{\sqrt{2\pi\omega\, r^{n+1}}}\Big( (C_{1}-iC_{2})e^{i\omega r} + (C_{1}+iC_{2})e^{-i\omega r} \Big)
\end{eqnarray} 
%
we infer the incoming amplitude and the incoming flux from infinity,
%
\begin{eqnarray}
F_{\text{in}} \,=\,\ \frac{ir^{n+1}}{2}  \left( \psi^{*}_{\text{in}}\frac{d}{dr}\psi_{\text{in}} -\psi_{\text{in}}\frac{d}{dr}\psi^{*}_{\text{in}} \right) \,=\,\frac{1}{2\pi}\Big| C_{1}+iC_{2} \Big|^{2}.
\end{eqnarray} 
%
Quasinormal modes would be obtained under the condition $C_{1}+iC_{2}=0$.

In the overlap region, and to leading order at large $n$, this solution gives (see appendix~\ref{app:overlap}) 
%
%
\beq
\psi(r)\to 
\begin{cases}
\displaystyle\frac{\sR^{-\frac{1+q_{+}+q_{-}}{2}}}{\sqrt{2\pi n\omega_c \tanh{\alpha_0}}}\lp K_{\hat\omega} C_{1}\, \sR^{q_{+}}
-\frac{2C_{2}}{K_{\hat\omega}}\, \sR^{q_{-}}  \rp, &\quad\hat\omega<\omega_c,\\ \\
\displaystyle\frac{\sR^{-\frac{1+q_{+}+q_{-}}{2}}}{\sqrt{2\pi n\omega_c \tan{\beta_0}}}\lp (C_1-iC_2){K_{\hat\omega}}\, \sR^{q_{+}}
+\frac{C_1+iC_2}{K_{\hat\omega}}\, \sR^{q_{-}} \rp,&\quad\hat\omega>\omega_c,
\end{cases}
 \label{farexp}
\eeq
with
%
\begin{eqnarray}
K_{\hat\omega}=
\begin{cases}
e^{-n\omega_c(\alpha_0-\tanh\alpha_0)}&\quad\hat\omega<\omega_c\,,\\
e^{-in\omega_c(\beta_0-\tan\beta_0)-i\pi/4}&\quad\hat\omega>\omega_c\,,
\end{cases}
\end{eqnarray} 
%
and $\alpha_0$ and $\beta_0$ defined by
\begin{equation}
\frac{\hat{\omega}}{\omega_c}=
\begin{cases}
\,\text{sech}\,\alpha_0\,,&\hat\omega<\omega_c\,,\\
\,\text{sec}\,\beta_0\,,&\omega_c<\hat\omega\,,
\end{cases}
\end{equation} 
so that
\beq 
q_{+}-q_{-}=2\sqrt{\omega_c^2-\hat\omega^2}=2\omega_c\tanh \alpha_0=2i\omega_c\tan \beta_0\,.
\eeq

\subsection{Matching and analysis of results}\label{subsec:scalarmatch}

Matching the coefficients in eqs.~(\ref{nearexp}) and (\ref{farexp}) we find
%
\begin{eqnarray}
C_{1}&=&\frac{\sqrt{\pi(q_{+}-q_{-})}\,\Gamma(q_{+}+q_{-})\,\Gamma(q_{+}-q_{-})}{K_{\hat\omega}\Gamma (q_{+})^{2}}, \notag \\
C_{2}&=&-\frac{K_{\hat\omega}\sqrt{\pi (q_{+}-q_{-})}\,\Gamma(q_{+}+q_{-})\,\Gamma(q_{-}-q_{+})}{2\Gamma(q_{-})^{2}}, \label{c1c2}
\end{eqnarray} 
%
when $\hat\omega<\omega_c$, and
%
\begin{eqnarray}
C_{1}-iC_{2}&=&\frac{\sqrt{i\pi(q_{+}-q_{-})}\,\Gamma(q_{+}+q_{-})\,\Gamma(q_{+}-q_{-})}{K_{\hat\omega}\,\Gamma (q_{+})^{2}}, \notag \\
C_{1}+iC_{2}&=&\frac{K_{\hat\omega}\sqrt{i\pi (q_{+}-q_{-})}\,\Gamma(q_{+}+q_{-})\,\Gamma(q_{-}-q_{+})}{\Gamma(q_{-})^{2}}. \label{c1c22}
\end{eqnarray}
%
when $\hat\omega>\omega_c$.

This solves the problem of scalar field propagation in the presence of the black hole, since using these results we obtain the reflection and transmission amplitudes
\beq
R_l(\omega)=\frac{C_1-iC_2}{C_1+i C_2}\,,\qquad 
T_l(\omega)=\frac{\sqrt{2\pi\hat\omega}}{C_1+iC_2}\,,\label{RTom}
\eeq
which satisfy $|R_l|^2+|T_l|^2=1$.
 A simple quantity of interest is the absorption probability,
%
\begin{eqnarray}
\gamma_l(\omega)\,= \,\frac{F_{\text{horizon}}}{F_{\text{in}}}\,=\,\frac{2\pi\hat\omega }{\big|C_{1}+iC_{2}\big|^{2}} .
\end{eqnarray} 
%

Observe that all the dependence on $n$ in $C_1$ and $C_2$ is contained in the factor $K_{\hat\omega}$.

\subsubsection{Low frequency: $\hat\omega<\omega_c$}

Since $\alpha_0-\tanh\alpha_0 >0$, in this regime $K_{\hat\omega}$ is exponentially small in $n$, 
except for $\omega_c-\hat\omega\sim n^{-2/3}$ where $K_{\hat\omega}$ rapidly approaches $1$. 
We can write
\begin{eqnarray}
K_{\hat\omega<\omega_c}=e^{n\sqrt{\omega_c^2-\hat\omega^2}}\left(\frac{\omega_c+\sqrt{\omega_c^2-\hat\omega^2}}{\hat{\omega}} \right)^{-n\omega_c}\,.
\end{eqnarray} 

Given that $K_{\hat\omega}\ll 1$, the amplitude is dominated by $|C_{1}| \gg |C_{2}|$ and we approximate
%
\begin{eqnarray}\label{lowgamma}
\gamma_l(\omega)&\simeq& \frac{2\pi\hat\omega}{|{C_{1}}|^{2}}\notag\\
&=&\frac{\hat{\omega}K_{\hat\omega}^{2}}{\sqrt{\omega_c^2-\hat\omega^2}\,\Gamma(q_{+}-q_{-})^{2}}\frac{| \Gamma(q_{+})|^{4}}{|\Gamma(1+i2\hat{\omega})|^{2}},
\end{eqnarray} 
which is strongly suppressed.
More explicitly, if $\hat\omega\ll \omega_c$, and since $\omega_c\geq 1/2$, we can write
\beq\label{Kom2}
K_{\hat\omega}^2\simeq \lp\frac{e\hat\omega}{2\omega_c}\rp^{2n\omega_c}\ll 1\,,
\eeq
with all the other factors in \eqref{lowgamma} remaining of order one. Thus, restoring the radius $r_0$ and $\omega=n\hat\omega$ for clearer illustration, we conclude that waves of frequencies $\omega\lesssim n/r_0$ are very strongly reflected by the black hole and interact very little with it.

At very low frequencies $\omega r_0\ll 1$ we can check against earlier results.
Consider s-waves, $l=0$, \ie\ $\omega_c=1/2$, which are the dominant component of the absorption. In this case we have
\beq
\gamma_l(\omega)\simeq \frac{2\omega r_0}{n}K_{\hat\omega}^2
\eeq
with
%
\begin{eqnarray}\label{K2}
K_{\hat\omega}^2\,&\simeq&\,\left(\frac{e\omega r_0}{n}\right)^{n}
\lp 1-\frac{\omega^2 r_0^2}{n}+O(\omega^3, n^{-2})\rp\\
&\simeq&\,
\left(\frac{\omega r_0}{2}\right)^{n}\frac{n\pi}{\Gamma((n+2)/2)^2}
\lp 1-\frac{\omega^2 r_0^2}{n}+O(\omega^3, n^{-2})\rp\,,
\end{eqnarray} 
%
where we have used Stirling's formula. Introducing the horizon area $A_H=r_0^{n+1}\Omega_{n+1}$ we find
%
\begin{eqnarray}
\gamma(\omega)_{l=0}\,=\, \frac{\omega^{n+1}\Omega_{n+1}}{(2\pi)^{n+1}}A_{H}
\lp 1-\frac{\omega^2 r_0^2}{n}+O(\omega^3, n^{-2})\rp\,.
\end{eqnarray} 
%
Finally projecting plane waves onto s-waves, we find the scalar absorption cross section
\beq\label{sigmas}
\sigma^{\text{s-wave}}_{\text{abs}}=A_{H}\lp 1-\frac{\omega^2 r_0^2}{n}+O(\omega^3, n^{-2})\rp\,.
\eeq
The leading term is the universal result of \cite{Das:1996we}.

\subsubsection{High frequency: $\hat\omega>\omega_c$}

In this case $K_{\hat\omega>\omega_c}$ is purely imaginary, with
\beq
|K_{\hat\omega>\omega_c}|=1\,.
\eeq
When $\hat\omega\gg\omega_c$ it can be approximated by $K_{\hat\omega>\omega_c}\simeq e^{-i n\hat\omega}$.

In contrast to the previous regime, the absorption probability now does not depend on $K_{\hat\omega}$. It takes the simple form
\beq\label{gammahigh}
\gamma_l(\hat\omega)=\frac{\sinh\lp 2\pi\hat\omega\rp\, \sinh \lp 2\pi\sqrt{\hat\omega^2-\omega_c^2}\rp}
{\left( \cosh\left[\pi\lp\hat\omega+\sqrt{\hat\omega^2-\omega_c^2}\rp\right]\right)^2}\,.
\eeq
When $\hat\omega\gg \omega_c$ this becomes
\beq
\gamma_l(\omega)=1-O\lp e^{-2\pi\hat\omega}\rp\,,
\eeq
which is the expected result: the black hole is an almost perfect absorber at very high frequencies. 

Note that the absence of a dependence on $n$ in \eqref{gammahigh} implies a scaling behavior with $\omega$ and $l$ of the absorption probability at large $n$.

\subsection{Effective scalar wave dynamics}\label{subsec:effdyn}

Matching constructions like we have performed admit an interpretation as effective theories. In our case case, the theory consists of scalar waves that propagate in flat space. When they reach a horizon (at $r=1$) the field modes $\psi_{\hat\omega\hat l}(r)$ must satisfy the boundary condition (from \eqref{nearexp}) that
\beqa\label{effbdry}
\left.\frac{\partial_r\psi_{\hat\omega\hat l}}{\psi_{\hat\omega\hat l}}\right|_{r=1}
=\frac{n}{2}\lp
(q_+-q_-)\frac{\Gamma(q_{+}-q_{-})\Gamma(q_{-})^{2}-\Gamma(q_{-}-q_{+})\Gamma(q_{+})^{2}}{\Gamma(q_{+}-q_{-})\Gamma(q_{-})^{2}+\Gamma(q_{-}-q_{+})\Gamma(q_{+})^{2}}-1
\rp\,.
\eeqa
This equation \textit{encodes all the scalar dynamics of the black hole}, to leading order at large $n$. Once this condition is imposed, the reflection and transmission amplitudes for field propagation take the form that we determined above. 

This effective description differs in important respects from the one that results from the more familiar study of black hole absorption at very low frequency $\omega\ll 1/r_0$ \cite{Starobinsky:1973,Unruh:1976fm,Harmark:2007jy}. That analysis also performs a matched asymptotic expansion between a near region $r\ll \omega^{-1}$ and a far region $r\gg r_0$. 
Then the effective theory is obtained after integrating out the degrees of freedom at scales $<r_0$. In the large $n$ effective theory, instead, we integrate the physics near the horizon at scales $<r_0/D$.

This distinction is important. The conventional low-frequency effective theory treats the black hole as a point particle: waves with $\omega\ll 1/r_0$ do not resolve the size of the horizon. The conditions on the far field are then effectively imposed at $r=0$. 
In our large $n$ effective theory, instead, the horizon radius $r_0=1$ remains finite (even if the horizon area is governed by  a much smaller scale when $n\to\infty$), so the boundary conditions are imposed on a sphere of finite radius. This is why we are able to capture a much larger frequency range, including $\omega> n/r_0\gg 1/r_0$, in which the black hole acts as an almost perfect absorber: excitations of these wavelengths perceive it as a finite-size object.

\subsection{Final remarks}

\paragraph{Upside.} We have obtained a solution for the scattering and absorption problem, eqs.~(\ref{c1c2}, \ref{c1c22}, \ref{RTom}), which gives in a simple analytic expression the amplitudes over a very wide range of frequencies (and partial waves), running from the lowest part of the spectrum until very high values. As an illustration, we not only recover the well-known universal result for the low-energy absorption cross section but also we can easily extract corrections to it, eq.~\eqref{sigmas}, which would be very hard to obtain in other approaches.

The results also confirm the conclusions of section~\ref{sec:radn}: the interaction of the black hole with waves of frequencies that remain fixed as $D$ grows is strongly suppressed by a factor of the form $\sim (\omega r_0/D)^D$ (see \eg\ eq.~\eqref{Kom2} or \eqref{K2}), while frequencies that scale like $D$ interact appreciably with the black hole. In fact waves with 
\beq
\omega\gg \frac{D}{r_0}\lp \frac12+\frac{l}{D}\rp
\eeq
are almost perfectly absorbed.

Our analysis also provides the solution for several types of linearized gravitational perturbations of this black hole \cite{Kodama:2003jz}: perturbations that are tensors on $S^{D-2}$ are governed by the same equation as \eqref{psieq} at all $D$, and scalar gravitational perturbations also obey, to leading order at large $D$, the equations \eqref{psieq3}, \eqref{hatpot} when $l=O(D^0)$. Moreover, gauge field perturbations satisfy the same equations at large $D$ including $\hat l=O(D^0)$. Shear perturbations of black branes are also known to obey the equations for massless scalars.

\paragraph{Downside.}

During the matching construction we have found that it breaks down in some specific instances.

The first one occurs when $q_+\simeq q_-$, \ie\ when $\hat\omega\simeq \omega_c$. This is also the region in which $|C_1/C_2|\simeq 1$. Since the quasinormal modes appear when $C_1+iC_2=0$, we see that our present construction is not well suited for the calculation of their frequencies. Nevertheless, this analysis shows that quasinormal modes must be expected when $\hat\omega\sim \omega_c$. This agrees with the WKB considerations in sec.~\ref{sec:scalarwave} and refs.~\cite{Konoplya:2003ii,Berti:2003si,Berti:2009kk}. Quasinormal modes are of course very important since they are responsible for resonant scattering which results in damped ringing at their frequencies. We expect that a modification of our technique will allow an analytic solution of the quasinormal modes and their frequencies. 

Second, our matching is also incorrectly performed when $q_{+}-q_{-}\in \mathbb{N}$. While the matching result for $C_{1}$ may still be correct, the matching of $C_{2}$ is more complicated. 

Our solution should lose accuracy around these special frequencies. For instance, at all of them the absorption appears to vanish since $C_2$ diverges, but this conclusion is not reliable. 
In constrast, for all $\hat\omega>\omega_c$ there is no problem in matching $C_{1}$ and $C_{2}$ and our results in this range away from $\hat\omega\sim \omega_c$ are sound.

Finally, the effective theory approach fails to apply for the ultra-high frequencies and angular momenta in the range $\omega r_0,l\gtrsim D^2$. The wavelength of these excitations is so short that they are insensitive to the curvature and they do not distinguish between near and far regions. Still, our expression for the absorption probability can be expected to smoothly merge with the result at these ultra-high frequencies when $\omega r_0\gtrsim D^2,l$, since in this case the deviations from $\gamma_l(\omega)=1$ must be extremely small. However, the range $\omega r_0\sim l\gg D^2$ seems to lie outside the large $D$ techniques of this section.

\paragraph{Aside: near-horizon conformal symmetry?} 

The appearance of hypergeometric functions in the near-region solution is suggestive of the presence of a two-dimensional conformal symmetry governing the amplitudes \cite{Maldacena:1996ix,Castro:2010fd}. Although in this article we will not pursue this idea, let us mention a potentially relevant fact that arises when $q_{+}-q_{-}\in \mathbb{N}$,\footnote{In these cases the matching for $C_{2}$ is not valid, but we expect that the result for $C_{1}$, which is what we use, remains valid.} which is always in the regime $\hat\omega<\omega_c$.
For odd $q_{+}-q_{-}=2m+1$, the absorption probability is
%
\begin{eqnarray}
\gamma^{\text{odd}}(\omega)\,=\,\frac{2\hat{\omega}K_{\hat\omega}^{2}}{ \Gamma(2m+1)\Gamma(2m+2)}
\frac{\Big| \Gamma(1+m+ i\frac{\omega}{4\pi T_{H}})\Big|^{4}}{\Big| \Gamma(1+i\frac{\omega}{2\pi T_{H}}) \Big|^{2}},
\end{eqnarray} 
%
and for even $q_{+}-q_{-}=2m$,
%
\begin{eqnarray}
\gamma^{\text{even}}(\omega)\,=\,\frac{2\hat{\omega}K_{\hat\omega}^{2}}{ \Gamma(2m)\Gamma(2m+1)}
\frac{\Big| \Gamma(\frac{1}{2}+m+ i\frac{\omega}{4\pi T_{H}})\Big|^{4}}{\Big| \Gamma(1+i\frac{\omega}{2\pi T_{H}}) \Big|^{2}}. 
\end{eqnarray} 
%
Using
%
\begin{eqnarray}
\Big|\Gamma(1+m+ix)\Big|^{2}&=&\frac{2\pi x e^{-\pi x}}{1-e^{-2\pi x}}\prod^{m}_{s=1}(s^{2}+x^{2})\,, \notag \\
\Big|\Gamma(\frac12+m+ix)\Big|^{2}&=&\frac{2\pi e^{-\pi x}}{1+e^{-2\pi x}}\prod^{m}_{s=1}(s^{2}+x^{2}), 
\end{eqnarray} 
%
the absorption factors can be written as bosonic or fermionic Boltzmann thermal functions at temperature $T_H$ (or two sectors at $T_H/2$). 

The conformal symmetry in \cite{Castro:2010fd} appears in the low frequency regime $\omega r_0\ll 1$. As discussed in sec.~\ref{subsec:effdyn}, our approach can deal with much larger frequencies, capturing the dynamics up to $\omega<D^2/r_0$. Then the amplitudes and absorption probabilities above, although similar to the ones in \cite{Castro:2010fd}, differ from them. It would be very interesting if the large $D$ effective description for the black hole took the the form of a two-dimensional conformal theory.

\section{Black brane instability}
\label{sec:GL}

In the previous section we have solved a scattering problem, namely the interaction of the black hole with waves that propagate outside it. In this section we investigate a problem closer in spirit to quasinormal mode analysis, where we study the dynamics of the black hole itself. This is, we solve an eigenvalue problem for the field in the near region with boundary conditions that correspond to the absence of any external perturbing sources outside the black hole. The modes we seek differ from quasinormal ones in that they correspond to an instability.

Refs.~\cite{Gregory:1993vy,Gregory:1994bj} proved, through a numerical solution of linearized perturbation equations, that the black branes of \eqref{bbrane} are unstable to fluctuations along their worldvolume, $\delta g_{\mu\nu}\sim e^{\Omega t+i \mathbf{ k\cdot z}}$. Besides demonstrating the large $D$ expansion in a different problem, the determination of the spectrum $\Omega(k)$ is an excellent benchmark for large $D$ studies: good numerical results are available in several dimensions, but also there exist analytical approximate results \cite{Kol:2004pn,Asnin:2007rw,Camps:2010br,Caldarelli:2012hy}. Thus we can test both the accuracy of the method when applied to finite values of $D$, and its effectiveness in comparison to other approaches. The outcome is very good on both counts.

\subsection{Perturbation equations}

At linearized order the perturbation problem depends on 
\beq
n=D-p-3
\eeq 
but not on $D$ and $p$ separately, so in this section we will use $n$ as the expansion parameter. To lighten the notation we set $r_0=1$ like in the previous section.

The Gregory-Laflamme problem involves perturbations that are scalars on $S^{n+1}$. There is only one physical degree of freedom, which by an appropriate choice of gauge can be chosen to be
\beq
\delta g_{tr}=\eta(r)e^{\Omega t+i kz}\,,
\eeq
all other metric components being obtainable from this one. The linearized perturbation equation is
\beq\label{GLeq}
\eta''+P(r) \eta'+Q(r)\eta=0
\eeq
where the functions $P(r)$ and $Q(r)$ are \cite{Gregory:1994bj}
\begin{eqnarray}
P(r)=\frac{1}{rfA(r)}
&\Big[& 3n^{3}-12n\Omega^{2}r^{2}+(3n^{2}-6n^{3}-8nk^{2}r^{2}-4r^{2}\Omega^{2}+8n\Omega^{2}r^{2})f \notag \\
&&
-(6n^{2}-3n^{3}+4k^{2}r^{2}-4nk^{2}r^{2})f^{2} +3n^{2}f^{3} \Big] \,,
\end{eqnarray}
\begin{eqnarray}
Q(r)=\frac{1}{r^{2}f^{2}A(r)}
&\Big[& (n^2-\Omega^2 r^2)(n^2-4\Omega^2 r^2)\notag \\
&+&(3n^{3}-n^{4}+n^{2}k^{2}r^{2}+10n^{2}\Omega^{2}r^{2}
+8k^{2}\Omega^{2}r^{4})f \notag \\
&+&\lp n^2(1-6n-n^2)+2k^2r^2(2k^2r^2-2n-n^2)+\Omega^2r^2(4+4n-5n^2)
\rp f^2\notag \\
& +&(-2n^{2}+3n^{3}+n^{4}+4k^{2}r^{2} 
+8nk^{2}r^{2}+n^{2}k^{2}r^{2})f^{3}+ n^{2}f^{4}\Big],
\end{eqnarray}
%
with
%
\begin{eqnarray}
A(r)\,=\,n^{2}-4\Omega^{2}r^{2}-(4k^{2}r^{2}+2n^{2})f+n^{2}f^{2}.
\end{eqnarray}
%

We shall solve the boundary value problem that results from requiring asymptotic flatness at infinity and regularity at the horizon. 
Note that $P$ and $Q$ have a simple pole at $r=r_{s}>1$ where $A(r_{s})=0$. This is a regular singular point of the equation. We have analyzed this point and checked that the solution that we find below is indeed regular there.

Taking cue from the result \eqref{GLD} that the zero-mode wavenumber scales like $1/\sqrt{n}$, we define
\beq\label{khatk}
\hat k =\frac{k}{\sqrt{n}}\,,
\eeq
and regard $\hat k$ as being $O(n^0)$, which allows us to keep track of very short wavelengths $\sim r_0/\sqrt{n}$. In contrast we take $\Omega=O(n^0)$. 

We solve the problem by matching the solutions in the near and far regions of sec.~\ref{subsec:nearfar}. Our method is an adaptation and extension of ref.~\cite{Asnin:2007rw}.

\subsection{Far region}\label{subsec:farGL}

In this region $r^{-n}$ is a quantity that is exponentially small in $1/n$ and therefore does not yield any perturbative $1/n$ corrections. Eq.~(\ref{GLeq}) for $\eta^{(\text{far})}$ becomes the flat space equation
%
\begin{eqnarray}\label{fareqn}
\frac{ d^{2}\eta^{(\text{far})}}{dr^{2}} +\frac{n+1}{r}\frac{d \eta^{(\text{far})}}{dr} 
-\lp \frac{n+1}{r^2}+n\hat{k}^{2}+\Omega^{2}\rp\eta^{(\text{far})}\,=\,0.
\end{eqnarray} 
%
Introducing the notations $\nu=(n+2)/2$ and $k_{\Omega}=\sqrt{n\hat{k}^{2}+\Omega^{2}}$, the solution that is regular at infinity is
%
\beq
\eta^{(\text{far})}\,\propto\frac{K_{\nu}(k_{\Omega} r)}{r^{n/2}}, \label{farsoln}
\eeq 
%
where $K_{\nu}(x)$ is the modified Bessel function of the second kind. 
We fix the integration constant for the amplitude of the perturbation in a manner that will simplify the matching in the overlap region. Here we use the coordinate $\sR$ defined in \eqref{sR} and expand up to next-to-next-to-leading order,
\begin{eqnarray}
r\,=\,1+\frac{\ln{\sR}}{n}+\frac{(\ln{\sR})^{2}}{2n^{2}}+O(n^{-3})\,.
\end{eqnarray}
Expanding now (\ref{farsoln}) for large $n$ we find
%
\begin{eqnarray}
\frac{K_{\nu}(k_{\Omega} r)}{r^{n/2}} 
&=&A_n\Biggl(\frac{1}{\sR}-\frac{1+\hat{k}^{2}}{n}\frac{\ln{\sR}}{\sR} \notag \\
&&\quad\qquad+\frac{1}{n^{2}}\frac{2(\hat{k}^{4}-\Omega^{2})\ln{\sR}+(1+\hat{k}^{4})(\ln{\sR})^{2}}{2\sR} +O(n^{-3})  \Biggr) \,.
\label{Kexpand}
\end{eqnarray}
%
Here $A_n$ is a factor (independent of $\sR$) in which we absorb all the $n$-dependence from terms of the form $\propto 1/(n^{i}\sR)$ in this expansion. We choose the integration constant in $\eta^{(\text{far})}$ to eliminate this factor, so that
%
\begin{eqnarray}
\eta^{(\text{far})} 
&=&\frac{1}{\sR}-\frac{1+\hat{k}^{2}}{n}\frac{\ln{\sR}}{\sR} \notag \\
&&
+\frac{1}{n^{2}}\frac{2(\hat{k}^{4}-\Omega^{2})\ln{\sR}+(1+\hat{k}^{4})(\ln{\sR})^{2}}{2\sR} +O(n^{-3}).   
\label{farOL}
\end{eqnarray}
Two comments about this result. First, note the absence of a constant term $\propto \sR^0$. This is a consequence of asymptotic flatness. The second independent solution to \eqref{fareqn} is $I_{\nu}(k_{\Omega} r)/r^{n/2}$, which grows at $r\to\infty$: this is a non-normalizable perturbation, which amounts to introducing sources for the field at infinity. This second solution would also yield a term $\propto \sR^0$ in the overlap region. Therefore the requirement that in this region
\beq\label{overcond}
\lim_{\sR\to\infty}\eta=0\,,
\eeq
is equivalent to asymptotic flatness in the far region.

Second, the expansion \eqref{farOL} involves higher powers of $\ln \sR$ but not of $1/\sR$ (this is true also if we include non-normalizable perturbations), essentially owing to the strong localization that makes $f=1+O(e^{-n})$. Nevertheless, it is possible to compute the $1/\sR^2$ terms that come from the next-order far solution. Although we will proceed without them, they may be used to provide an alternative matching calculation. In appendix~\ref{app:nextfar} we give some details of this.


\subsection{Near region}\label{subsec:nearreg}

We expand in $1/n$ as
\beq
\eta\,=\,\sum_{j\geq 0}\frac{\eta_{(j)}(\sR)}{n^{j}}.
\eeq
Eq.~(\ref{GLeq}) gives an equation at each perturbative order of the form
\beq\label{kth-eq}
\frac{d}{d\sR}\left(\sR(\sR-1)^{3}\frac{d\eta_{(j)}}{d\sR}\right) +(2\sR-1)(\sR-1)\eta_{(j)}\,=\,\mathcal{S}_{(j)}, 
\eeq
where $\mathcal{S}_{(j)}$ is a $j$-th order source term built out of the solution up to order $j-1$. 
The homogeneous equation has as its independent solutions
%
\beq
u_{0}=\frac{1}{\sR-1}\,,\qquad v_{0}=\frac{\ln{(\sR-1)}-\ln{\sR}}{\sR-1}.
\eeq
%
With these the solution to eq.~(\ref{kth-eq}) can be found using Green's method in the form
%
\beqa
\eta_{(j)}&=&A_{j}u_{0} +B_{j}v_{0} \notag \\
&&
+u_{0}(\sR)\int^{\infty}_{\sR}v_{0}(\sR')\mathcal{S}_{(j)}(\sR')d\sR' 
+v_{0}(\sR)\int^{\sR}_{1}u_{0}(\sR')\mathcal{S}_{(j)}(\sR')d\sR'. \label{greenfn}
\eeqa
%
The integration constants $A_{j}$ and $B_{j}$ will be determined by matching to 
the far region solution (\ref{farsoln}) at large $\sR$, and requiring regularity on the horizon at $\sR=1$. This last condition can be derived by solving eq.~\eqref{GLeq} near $r=1$. The regular solution is
%
\begin{eqnarray}
\eta &\propto&  (r-1)^{-1+\Omega/n} \lp 1+O(r-1)\rp\notag \\
&\propto& \frac{1}{\sR-1}\Biggl( 1+\frac{\Omega}{n}\ln{(\sR-1)} 
+\frac{\Omega^{2}}{2n^{2}} (\ln{(\sR-1)})^{2} \notag \\
&&\qquad\qquad
+\frac{\Omega^{3}}{6n^{3}} (\ln{(\sR-1)})^{3}+O(n^{-4},\sR-1) \Biggr). \label{bcnear}
\end{eqnarray} 
%

\paragraph{Matching condition.}

The overlap region corresponds in the near region to $\sR\gg 1$. Eq.~\eqref{overcond} then implies that 
%
\beq
\lim_{\sR\to\infty}\sum_j\frac{\eta_{(j)}}{n^j}=0\,. \label{regcond}
\eeq
%
This condition can easily be seen to imply that the Wronskians of $u_{0}$ and $\eta_{(j)}$,
%
\beq
W[u_{0},\eta_{(j)}]=\sR(\sR-1)^{3}[ u_{0}(\sR)\eta'_{(j)}(\sR)-u'_{0}(\sR)\eta_{(j)}(\sR) ]\,,
\eeq
%
satisfy
\beq
\lim_{\sR\rightarrow\infty}\frac{1}{\sR^{2}}\sum_{j}\frac{W[u_{0},\eta_{(j)}]}{n^{j}}=0\,.
\eeq
Since these Wronskians are given by
%
\beq
W[u_{0},\eta_{(j)}]=\int^{\sR}d\sR\, u_{0}\,\mathcal{S}_{(j)}\,, \label{wronskian}
\eeq
%
then the boundary condition in the overlap region can be conveniently written in the form
%
\beq
\lim_{\sR\rightarrow\infty}\frac{1}{\sR^{2}}\sum_{j}\frac{1}{n^{j}}\int^{\sR}d\sR\, u_{0}\,\mathcal{S}_{(j)} =0\,. \label{cond}
\eeq
%
Crucially, this $j$-th order boundary condition can be imposed with knowledge of the solution just up to $(j-1)$-th order.

\medskip

We can now proceed to solve \eqref{kth-eq} order by order, with the sources $\mathcal{S}_{(j)}$ computed in appendix~\ref{app:sources} and with the boundary conditions \eqref{bcnear} and \eqref{cond}.

\paragraph{Zeroth order.}

The equation at this order is homogeneous, so
%
\beq\label{eta0}
\eta_{(0)}\,=\,\frac{A_{0}}{\sR-1}+B_{0}\frac{\ln{(\sR-1)}-\ln{\sR}}{\sR-1}.
\eeq
%
Horizon regularity (\ref{bcnear}) fixes $B_{0}=0$. Matching to the far solution in the overlap region (\ref{farOL}) fixes the amplitude $A_{0}=1$.

\paragraph{First order.} 
The condition (\ref{cond}) for $\eta_{(1)}$ is automatically satisfied. We can integrate the 
source terms in \eqref{greenfn} and determine the constants $A_1$, $B_1$ by imposing the boundary condition at the horizon and matching with the far region solution. Thus we obtain the first order solution in the near region as
%
\beq\label{eta1}
\eta_{(1)}=\frac{\Omega\ln{(\sR-1)}-(1+\hat{k}^{2}+\Omega)\ln{\sR}}{\sR-1}.
\eeq
%

\paragraph{Second order.}

The condition (\ref{cond}) to this order requires that
%
\beq
-\frac{2}{n^{2}}(\Omega^{2}-\hat{k}^{2}+2\hat{k}^{2}\Omega +\hat{k}^{4}) =0. 
\eeq
Choosing $\hat k\geq 0$, the solution to this equation that gives unstable modes is
\beq\label{Omvsk0}
\Omega=\hat{k}-\hat{k}^{2}\,.
\eeq 
This gives the dispersion relation for the Gregory-Laflamme instability to leading order at large $n$.

As discussed above, this result does not require the actual second-order solution: only the source term computed with $\eta_{(1)}$ has been used. However, we will need $\eta_{(2)}$ in order to find further corrections. Integrating eq.~(\ref{kth-eq}) for $j=2$ we find
%
\begin{eqnarray}\label{eta2}
\eta_{(2)}&=&-\frac{\pi^{2}\Omega^{2}-2\pi^{2}\hat{k}^{2}+2\pi^{2}\Omega\hat{k}^{2}+\pi^{2}\hat{k}^{4}}{6(\sR-1)} \notag \\
&&
+\frac{1}{2(\sR-1)}\Bigl( \Omega^{2}(\ln{(\sR-1)})^{2} 
+(2\Omega +2\hat{k}^{2}+1)(\ln{\sR})^{2} \notag \\
&&\qquad\qquad\qquad
+2(\hat{k}^{4}-\Omega^{2}-\Omega(1+\hat{k}^{2}+\Omega)\ln{(\sR-1)})\ln{\sR} \notag \\
&&\qquad\qquad\qquad
-4(\hat{k}^{4}+\Omega^{2}+2\hat{k}^{2}\Omega-\hat{k}^{2})(\sR-1) \notag \\
&&\qquad\qquad\qquad
-2(\Omega^{2}+\hat{k}^{4}+2\hat{k}^{2}\Omega-2\hat{k}^{2})\mathrm{Li}_{2}(1-\sR) \Bigr)\,,
\end{eqnarray}
%
where $\mathrm{Li}_{2}(z)$ is the dilogarithm function.
The integration constants $A_{2}$ and $B_{2}$ have been fixed by matching to the far solution (\ref{farOL}) and to the horizon solution (\ref{bcnear}). In more detail, near the horizon the solution becomes
%
\begin{eqnarray}
&&\eta_{(0)}+\frac{1}{n}\eta_{(1)}+\frac{1}{n^{2}}\eta_{(2)} \Big|_{\sR=1} \notag \\
&&~~
=\left( 1+\frac{B}{n^{2}}\right)\Biggl( 1+\frac{\Omega}{n}\ln{(\sR-1)}+\frac{\Omega^{2}}{2n^{2}}(\ln{(\sR-1}))^{2}+O(\sR-1) 
\Biggr), 
\end{eqnarray}
%
where
%
\beq
B=\frac{-\pi^{2}\Omega^{2}+2\pi^{2}\hat{k}^{2}-2\pi^{2}\hat{k}^{2}\Omega -\pi^{2}\hat{k}^{4} }{6}. \label{B}
\eeq
%

\paragraph{Third order.}

The regularity condition (\ref{cond}) for $\eta_{(3)}$ becomes 
%
\begin{eqnarray}
&&-\frac{2}{n^{2}}(\Omega^{2}-\hat{k}^{2}+2\hat{k}^{2}\Omega +\hat{k}^{4}) \notag \\
&&~~~~
-\frac{2}{n^{3}}\Big[ -\Omega^{2} +2\Omega^{3} +2\hat{k}^{2} -2\hat{k}^{2}\Omega +2\hat{k}^{2}\Omega^{2} +\hat{k}^{4} 
-2\hat{k}^{4}{\Omega} -2\hat{k}^{6} \notag \\
&&~~~~~~~~~~~
+(1+\hat{k}^{2})(\Omega^{2}-\hat{k}^{2}+2\hat{k}^{2}\Omega +\hat{k}^{4})\ln{\sR} \Big]  \notag \\
&&~~
=O(n^{-4})\,,
\end{eqnarray}
%
whose solution
%
\beq\label{Omvsk1}
\Omega\,=\,\hat{k}-\hat{k}^{2} -\frac{\hat{k}}{2n}(1+2\hat{k}-2\hat{k}^{2}). 
\eeq
%
gives the dispersion relation to next-to-leading order in $1/n$. 
Again, this has been obtained before calculating $\eta_{(3)}$  from \eqref{greenfn}. 

The boundary condition at the horizon is satisfied by setting $B_{3}=B\Omega$ in (\ref{B}).
The second integral in \eqref{greenfn} can be performed analytically. This implies that we can compute the behavior of the third order solution $\eta_{(3)}$ in the overlap region explicitly, which is crucial in order to impose the regularity condition  (see appendix~\ref{app:sources}).
The constant $A_{3}$ is determined by matching to the far region solution (\ref{farOL}). Although this can be done explicitly, it turns out that $A_{3}$ does not enter in the condition \eqref{cond} for the fourth order solution. So we can proceed to the next step without specifying  $A_{3}$.

\paragraph{Fourth order.}

With the previous solution we can compute the source $\mathcal{S}_{(4)}$.
Condition \eqref{cond} at fourth order gives
%
\begin{eqnarray}
\label{Omvsk}
\Omega &=&\hat{k}-\hat{k}^{2} -\frac{\hat{k}}{2n}(1+2\hat{k}-2\hat{k}^{2}) \notag \\
&&
+\frac{\hat{k}}{24n^{2}}(9+24\hat{k} +12\hat{k}^{2}-8\pi^{2}\hat{k}^{2}+8\pi^{2}\hat{k}^{3}-12\hat{k}^{4}).
\end{eqnarray}
%

In order to proceed beyond this point we should do the first integral in eq.~(\ref{greenfn}) to impose the boundary condition on $\eta_{(4)}$ at the horizon. However, we have not managed to do this analytically. Then we cannot determine $B_{4}$, which affects the fifth order regularity condition that would give the $1/n^3$ term in $\Omega(\hat k)$. So we stop at this order. 

Eq.~\eqref{Omvsk} is the main result of this section.

\subsection{Comparisons and accuracy}

Setting $\Omega=0$ in \eqref{Omvsk} gives the wavenumber $\hat k$ of the threshold zero-mode. Reverting to $k=\sqrt{n}\hat k$, we find
%
\begin{eqnarray}
k_{\text{GL}}=\sqrt{n}\left( 1-\frac{1}{2n}+\frac{7}{8n^{2}} +O(n^{-3}) \right). 
\end{eqnarray}
%
This reproduces the result in \cite{Asnin:2007rw}, which was obtained with a method essentially similar to ours, but using a different gauge. We have discussed the interpretation of the leading order result earlier in \eqref{GLD}.

Analytic approximations to $\Omega(k)$ have been computed from a rather different approach. Refs.~\cite{Camps:2010br,Caldarelli:2012hy} solved black brane perturbations in a hydrodynamic expansion at small $k$ for arbitrary $n$. Ref.~\cite{Camps:2010br} conjectured that the relation $\Omega=\hat k-\hat k^2$ is exact when $n\to\infty$.
Our results, already in \eqref{Omvsk0}, do prove it.
Ref.~\cite{Caldarelli:2012hy} extended the calculation to include terms up to $\propto k^3$.
These hydrodynamic results and our large $n$ expansion agree where they overlap:
eq.~\eqref{Omvsk} to order $(n^{-2},\hat{k}^{3})$ is the same as the expansion of the result of \cite{Camps:2010br,Caldarelli:2012hy} to the same order.\footnote{See appendix~\ref{app:hydrovsn} for some additional comparison.} 

\begin{figure}[t]
\begin{center}
\includegraphics[width=.9\textwidth]{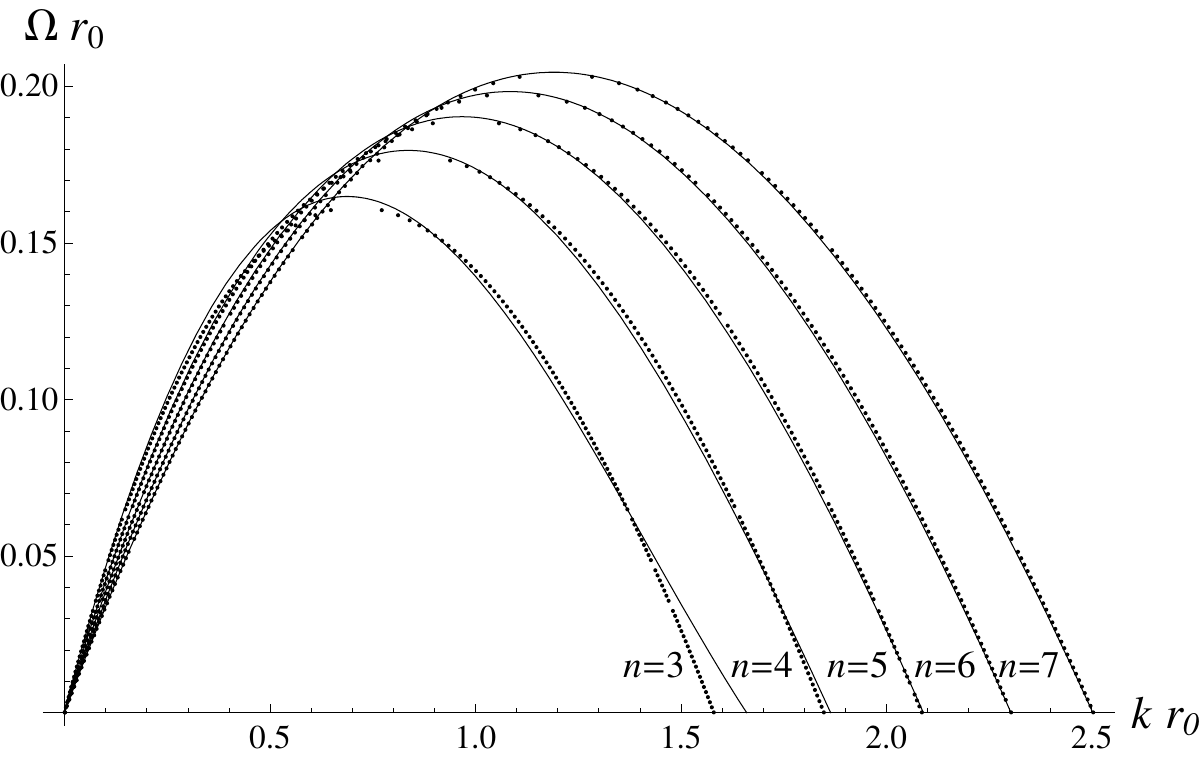} 
\end{center}
\caption{Dispersion
relation $\Omega(k)$ of unstable modes for $n=3,4,5,6,7$: the solid line is
our analytic approximation eq.~\eqref{Omvsk}; the dots are the
numerical solution (the same as in \cite{Camps:2010br}, courtesy of P.~Figueras).} 
\label{fig:Omvsk}
\end{figure}

\begin{figure}[t]
\begin{center}
\includegraphics[width=\textwidth]{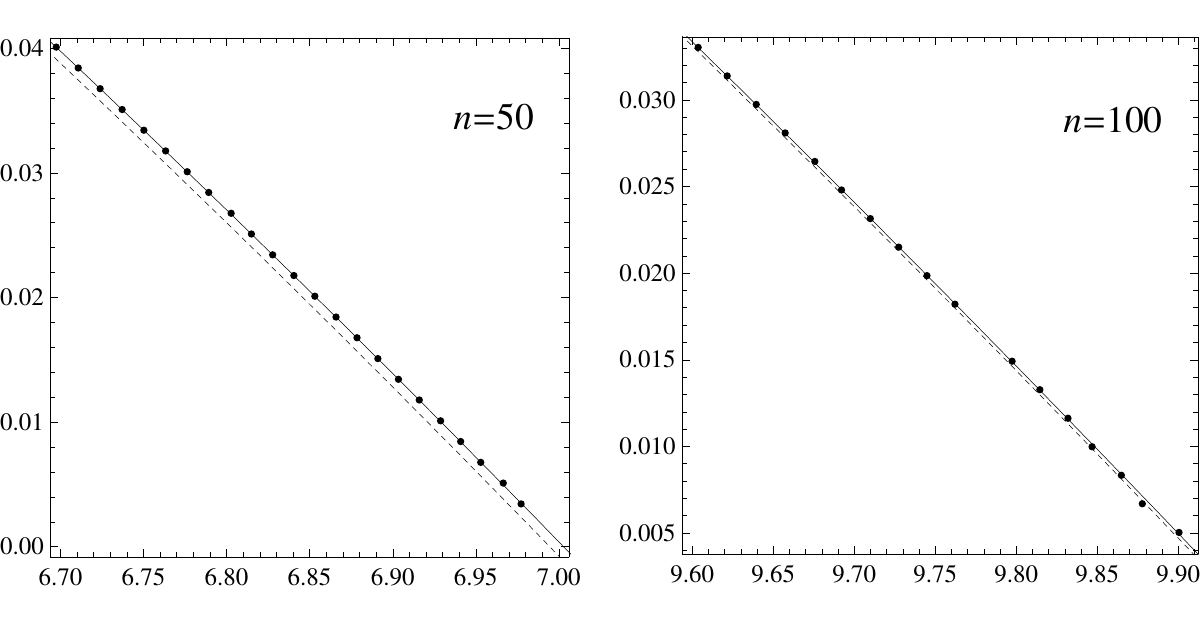} 
\end{center}
\caption{Dispersion
relation $\Omega(k)$ for $n=50,100$ for $k$ near the threshold zero-mode. The solid line is
our analytic approximation eq.~\eqref{Omvsk}; the dots are the
numerical solution; the dashed line is the analytic approximation of \cite{Caldarelli:2012hy}.} 
\label{fig:n50100}
\end{figure}

Eq.~\eqref{Omvsk}, however, also contains terms $\propto \hat k^4\,,\hat k^5$ which are new. They should give a more accurate dispersion relation at values $\hat k\sim 1$, where the hydrodynamic methods become less precise.
This accuracy is apparent in figure~\ref{fig:Omvsk}. While eq.~\eqref{Omvsk} gives a rather poor approximation to the dispersion curves for $n=1,2$ (which is not too surprising), it gives a very good match already for $n=3$. Overall, eq.~\eqref{Omvsk} is a much better fit to the entire curves than the hydrodynamic approximation in \cite{Caldarelli:2012hy}. E.g., for $n=3$ the hydrodynamic curve (which is exact in $n$) is better only at relatively small $k$ ($\lesssim 0.25$).

The agreement between \eqref{Omvsk} and the numerical data is, for large portions of the curves, quite better than the expected error $\sim 1/n^3$. The largest deviations tend to appear near the threshold mode: for $n=3$ we find $k_{\text{GL}}^{(num)}/k_{\text{GL}}^{(analytic)}\approx 0.96$, which is within the $\sim 1/n^3$ margins. At $n= 4$ the match is indeed much better, $k_{\text{GL}}^{(num)}/k_{\text{GL}}^{(analytic)}\approx 0.99$.

The accuracy at much larger values of $n$ is at least as good as the precision of the numerical data we have $(\sim 10^{-5}-10^{-6})$. This is shown in fig.~\ref{fig:n50100}, where we focus on the region near the zero mode where the discrepancies are possibly largest. For reference we also include the analytical result from \cite{Caldarelli:2012hy}.

\bigskip

\subsection{Final remarks}\label{subsec:fingl}

\paragraph{AdS black branes at large $D$.}

Ref.~\cite{Caldarelli:2012hy} describes a map that relates the dynamics of vacuum black $p$-branes in $D=n+p+3$ dimensions to the dynamics of AdS black branes in $d+1$ dimensions, by taking $n\leftrightarrow -d$. So, by a suitable analytic continuation of $n$ into the region of large negative numbers, we get a map of the solution in this section into a solution for perturbations of AdS$_{d+1}$ black branes at large $d$. Under this map the unstable mode becomes a stable, damped, quasinormal mode.

This solution goes beyond the hydrodynamic regime by being exact in the wavenumber $\hat k=k/\sqrt{d}$ at each order in $1/d$. Thus, it contains information about transport coefficients at all hydrodynamic orders. In particular, since to the order $1/d^2$ that we have solved, the terms $k^4$ and $k^5$ are non-zero, this gives non-zero values for certain combinations of third and fourth order transport terms.

\paragraph{Scalings.}
The modes relevant in this section are localized in the sphere of influence at $r-r_0\lesssim r_0/D$. However, the scaling of their frequencies and wavenumbers
\beq\label{OmkD}
\Omega r_0 \sim O(D^0)\,,\qquad k r_0\sim O(D^{1/2}),
\eeq
is different than the scaling $\omega r_0\sim O(D)$ of the characteristic black hole frequencies discussed in secs.~\ref{sec:radn} and \ref{sec:scalar}. They are also the scalings of a specific set of quasinormal modes of AdS black branes by the mapping above. 

The appearance of the scale $r_0/\sqrt{D}$ in these black brane modes seems to be related to the fact that they have a hydrodynamic limit, \ie\ their frequency vanishes as the wavenumber $k\to 0$. Hydrodynamic sound modes on black branes propagate with a velocity
\beq
\sqrt{\frac{dP}{d\varepsilon}}\sim \frac1{\sqrt{D}}
\eeq
(see \eqref{Pep} and \eqref{Pepc}; sound speed is imaginary for the unstable branes). Thus, in the hydrodynamic leading order, the dispersion relation is $\Omega,\omega\sim k/\sqrt{D}$. This determines the scaling of $\Omega/k$ but does not yet imply \eqref{OmkD}. Consider now the quadratic terms in the hydrodynamic dispersion relation, which have the form $(\eta/s)k^2/T$, involving the viscosity to entropy ratio. If viscous effects are to contribute at the leading large-$D$ order, then this quadratic term must scale with $D$ 
in the same way as the linear one. Since $T\sim D/r_0$, this requires that
\beq
k\sim \frac{\sqrt{D}}{r_0} \frac{s}{\eta}\,,
\eeq
and given that $s/\eta$ takes a universal, $D$-independent value for black objects, we recover the scalings in \eqref{OmkD}. It is not obvious why viscous effects should remain finite as $D\to\infty$, but we see that that is equivalent to having \eqref{OmkD} hold. It is suggestive that the scale of the entropy density is $\ell_A=r_0/\sqrt{D}$, so this is also the natural scale for $\eta$.

\bigskip

To conclude, the results in this section are quite encouraging for applications of the large $D$ expansion --- both because the calculations can be carried out explicitly up to a fairly high order, and because of their excellent quantitative agreeement down to rather low dimensions.

\section{Conclusion}\label{sec:concl}

Any gravitational problem that can be formulated in an arbitrary number of dimensions is susceptible to study in a large $D$ expansion. Since some terms in the equations will drop out of the expansion, this is virtually guaranteed to yield some simplification. Problems that otherwise require numerical study may in this approach become analytically tractable.

The basic physical intuition for why and how this can be so, lies in the remarkably consistent picture we have found for General Relativity at large $D$. To sum it up:

\begin{itemize}

\item Black hole physics at large $D$ separates into two regimes, `far' and `near', with length scales parametrically different in $D$,
\beq
\ell_\mathrm{near}=\frac{\ell_\mathrm{far}}{D}\,.
\eeq
$\ell_\mathrm{far}$ is the scale of the horizon radius, \ie\ $\ell_\mathrm{far}=r_0$ for the Schwarzschild solution~\eqref{schwd}.

\subitem{\bf Far region dynamics} 
 consists of fields in flat Minkowski geometries with `holes' removed at the location of horizons with large curvatures $\sim D/\ell_\mathrm{far}$. The radii of these holes are $\sim \ell_\mathrm{far}$, but their area is parametrically much smaller, $\sim (\ell_\mathrm{far}/\sqrt{D})^{D-2}$. Black hole interactions for separations on the scale $\ell_\mathrm{far}$ are trivial. Waves of frequencies $\sim 1/\ell_\mathrm{far}$ encounter horizons as perfectly reflective walls.

\subitem{\bf Near region dynamics} occurs within a thin region that extends out to a distance $\ell_\mathrm{near}\sim r_0/D$ from the horizon. This is where the gravitational potential and all the non-trivial dynamics of the black hole resides. Two black holes interact when their separation is $\lesssim \ell_\mathrm{near}$. Black holes emit and absorb waves of frequencies and wavenumbers $\gtrsim \ell_\mathrm{near}^{-1}$. 

\item This large-$D$ dynamics can be adequately captured by an effective theory of fields propagating in flat spacetime, subject to specific boundary conditions on the horizons, like \eqref{effbdry} for massless scalar fields.

\end{itemize}

The area-length (or mass-length) scale $r_0/\sqrt{D}$ appears to play a rather less significant role than $r_0$ and $r_0/D$, possibly restricted to the hydrodynamic modes of black branes.

One may, perhaps, feel uneasy about regarding a quantity as basic as the number of dimensions as a tunable parameter of a theory. However, this is not different than in $SU(N)$ gauge theories, or actually in most theories with adjustable parameters. We are viewing the gravity theories at different values of $D$ as being \textit{one} uniparametric theory, and eq.~\eqref{schwd} as \textit{one} uniparametric solution of this theory. 

We have studied only one of the limits of the parameter space of the theory, $D\to\infty$, but it may also be interesting to study the opposite limit. Arguably this corresponds not to $D=0$ or $1$ but to $D=3$. At $D=3$, the theory \eqref{Rmunu} ceases to have both local dynamics and black hole solutions, but taking $D$ as a continuous parameter one still finds non-trivial effects for $D=3+\epsilon$. This idea was put forward in ref.~\cite{Asnin:2007rw} to solve the problem of the Schwarzschild negative mode in an expansion in $\epsilon$. It should be of interest to understand this expansion in more generality.\footnote{See \cite{Grumiller:2007wb} for an expansion in $D=2+\epsilon$.}

Our study of gravitational radiation has been rather cursory, since at $D\to\infty$ it largely decouples from black holes --- except in phenomena where the radiation probes scales extremely close to the horizon. Nevertheless, gravitational waves are a sector of the theory that deserves closer attention. Some features of classical gravitational radiation depend on the spacetime dimensionality, \eg\ on whether $D$ is even or odd, as is the case for the applicability of Huygens principle \cite{Ooguri:1985nv,Cardoso:2002pa}. Further work is required to clarify the importance of all such effects in the large $D$ expansion.

We have not considered the coupling of the gravitational field to other fields or matter systems, except for brief references to charge and the cosmological constant. The main issue when considering these additions is the choice of how the new parameters scale with $D$. What choice is most convenient will depend on the particular type of problem one is interested in, but often it is useful to keep fixed the length scale that characterizes the effects of the new fields on the geometry. These remarks apply as well to solutions with compactified dimensions. The size of the compact space may be fixed/grow/shrink as $D$ increases, each choice capturing different physics. 
In a similar vein,
in the context of higher-dimensional gravity it is natural to also include the class of Lovelock theories. Each of the new terms that these theories introduce comes with a new length parameter. Depending on how these are chosen to scale with $D$, different large $D$ limits will result. The fact that the Riemann curvature tends to strongly localize close to the horizon indicates that the far- and near-region picture should still apply at least in some situations (\eg\ the `normal branch' of Lovelock black holes).

Finally, another setting in which the large $D$ expansion might be useful is the study of cosmologies of four-dimensional braneworlds in a large $D$ bulk. Having fixed worldvolume dimension, the gravitational effect of the brane in the bulk will be strongly localized close to the brane, which may simplify some calculations.

\section*{Acknowledgments}

RE acknowledges useful discussions with the authors of ref.~\cite{Caldarelli:2012hy}, and is also grateful to the organizers and participants in the workshop ``Holography, gauge theory and black holes" at the Institute of Physics, Univ.\ Amsterdam, in December 2012 where this work was presented. We thank Stanley Deser for correspondence on a previous version of this article. RS is grateful to the Departament de F{\'\i}sica Fonamental at Universitat de Barcelona for hospitality during the initial stage of this project. 
RE was partially supported by MEC FPA2010-20807-C02-02, AGAUR 2009-SGR-168 and CPAN CSD2007-00042 Consolider-Ingenio 2010. RS was supported by the Grant-in-Aid for the Global COE Program ``The Next Generation of Physics, Spun from Universality and Emergence" from the Ministry of Education, Culture, Sports, Science and Technology
(MEXT) of Japan. KT was supported by a grant for research abroad by JSPS.

\addcontentsline{toc}{section}{Appendices}
\addtocontents{toc}{\protect\setcounter{tocdepth}{0}}
\appendix

\section{Elementary geometry at large $D$}
\label{app:elemgeo}

Consider $D$-dimensional Minkowski spacetime. Using
\beq
\int_0^\pi d\theta (\sin\theta)^k=\sqrt{\pi}\,\frac{\Gamma\lp\frac{k+1}{2}\rp}{\Gamma\lp\frac{k+2}{2}\rp}\,,
\eeq
one obtains
the area of the unit-radius sphere $S^{D-2}$
\beqa
A_\mathrm{{sph}}&=&\int_0^\pi d\theta_1\int_0^\pi d\theta_2\dots\int_0^\pi d\theta_{D-3}\int_0^{2\pi}d\phi\, (\sin\theta_1)^{D-3} (\sin\theta_2)^{D-4}\dots \sin\theta_{D-3}\nonumber\\
&=&\Omega_{D-2}\,
\eeqa
as given in \eqref{OmD2}. This
is not a monotonic function of $D$: it reaches a maximum at $D\simeq 8.257$, and then rapidly decreases as $D$ grows. For integer $D$ the maximum is at $D=8$, where $\Omega_6=16\pi^3/15\simeq 33$. One might be tempted to conclude that this non-monotonicity could imply qualitative differences between values of $D$ smaller or larger than $8$, and thus a potential inadequacy of the large $D$ expansion when applied to $D<8$. 

However, this result does not make clear what we are comparing the sphere to. A more careful analysis shows that in an appropriate sense spheres do become monotonically small as $D$ grows.
The area of the circumscribed cube that contains the unit-radius sphere and is tangent to it at its faces, is
\beq
A_\mathrm{{circube}}=(D-1)2^{D-1}\,.
\eeq
The ratio $A_\mathrm{sph}/A_\mathrm{circube}$ can now be seen to be a \textit{monotonically decreasing} function of $D\geq 2$, which vanishes as
\beq\label{acirc}
\frac{A_\mathrm{sph}}{A_\mathrm{circube}}\to \frac{\sqrt{2}}{\pi}\left(\frac{\pi e}{2D}\right)^{D/2}\to 0\,.
\eeq

A simple intuition for this behavior follows from considering that the length $L$ of the diagonal of a cuboid in $D$ spacetime dimensions is
\beq
L^2=x_1^2+\dots +x_{D-1}^2\,.
\eeq
Then, for a generic cuboid at large $D$, its diagonal length $L$ is much larger than any of its side lengths $x_i$. In particular, the length $L$ of the half-diagonal of the circumscribed cube around a unit-radius sphere is
\beq
L=\sqrt{D-1}\gg 1\,.
\eeq
Now a plain indication that the sphere is in fact semifactorially smaller than its circumscribing cube is that
\beq
\lp\frac{1}{L}\rp^{D-2}\to D^{-D/2}\,,
\eeq 
which reproduces the leading behavior in \eqref{acirc}.

If we consider the inscribed cube (with its vertices on the sphere) then 
\beq
A_\mathrm{incube}=(D-1)^{2-D/2}2^{D-1}\,,
\eeq
which becomes much smaller than the sphere
\beq
\frac{A_\mathrm{incube}}{A_\mathrm{sph}}\to \sqrt{e} \left(\frac{\pi e}{2}\right)^{-D/2}\to 0\,.
\eeq
The behavior is monotonic for $D\geq 2$, but the shrinking rate is slower than in \eqref{acirc}, since $A_\mathrm{incube}\sim D^{-D/2}$, as follows from the above argument about the diagonals.

Similar behavior is obtained by comparing the volumes $V$ inside these bodies, since for the unit-radius sphere and for the cube of side-length two we have
\beq
V=\frac{A}{D-1}\,.
\eeq
However, these volumes do not appear to be relevant for any purpose in this paper.

\section{Technical appendices to sections \ref{sec:scalar} and \ref{sec:GL}}

\subsection{Far scalar wave solution in the overlap region}\label{app:overlap}

In the overlap region we have $\ln\sR\ll n$ and so we can expand
\beq\label{rover}
r=1+\frac{1}{n}\ln \sR+O(n^{-2})\,.
\eeq
Since at large $n$ and fixed $r$ both the argument and the index of the Bessel functions in \eqref{farscsoln} are large numbers of the same order $\sim n$,\footnote{So, again, the expansion is not valid when $\hat\omega\gtrsim n$.} it is appropriate to use the Debye expansion. When $\hat\omega<\omega_c$ it gives 
%
\begin{equation}
\psi(r)\,=\,\frac{1}{\sqrt{2\pi\sR\,n\omega_c\tanh\alpha}} \lp C_{1} e^{-n\omega_c(\alpha-\tanh\alpha)}-2C_{2}e^{n\omega_c(\alpha-\tanh\alpha)}+O(n^{-1})\rp\,,
\end{equation} 
%
while at $\hat\omega>\omega_c$,
\beqa
\psi(r)&=&\frac{1}{\sqrt{2\pi\sR\,n\omega_c\tan\beta}} \Bigl(
(C_{1}-iC_{2})e^{-in\omega_c(\beta-\tan\beta)-i\pi/4}\notag\\
&&\qquad\qquad\qquad\qquad
+(C_{1}+iC_{2})e^{in\omega_c(\beta-\tan\beta)+i\pi/4}+O(n^{-1})\Bigr)\,.
\eeqa 
Here $\alpha$ and $\beta$ are defined by
%
\begin{equation}
\frac{\hat{\omega}}{\omega_c}r=
\begin{cases}
\,\text{sech}\,\alpha\,,&\hat\omega<\omega_c\,,\\
\,\text{sec}\,\beta\,,&\omega_c<\hat\omega\,.
\end{cases}
\end{equation} 
%

Expanding in the overlap region using \eqref{rover} we find
%
\begin{eqnarray}
\alpha-\tanh\alpha =\alpha_0-\tanh\alpha_0
-\tanh \alpha_0\frac{\ln \sR}{n}+O(n^{-2}),
\end{eqnarray} 
%
and
\begin{eqnarray}
\beta-\tan\beta =\beta_0-\tan\beta_0
-\tan\beta_0\frac{\ln \sR}{n}+O(n^{-2})\,.
\end{eqnarray} 
With this, the far solution in this region can be written as in eq.~\eqref{farexp}.

\subsection{Black brane far solution at next order }\label{app:nextfar}

In the far region $r^{-n}$ is exponentially small in $n$ and therefore it does not yield any perturbative $1/n$ corrections. In order to compute corrections to the leading far solution \eqref{farsoln} we introduce an auxiliary order-counting parameter $\epsilon$, 
\beq
f(r)=1-\epsilon r^{-n}\,,
\eeq
which is set to $\epsilon=1$ at the end of the calculations. The far solution is then expanded in the form
\beq
\eta^{(far)}=\eta^{(0)}+\epsilon \eta^{(1)}+O(\epsilon^2).
\eeq

The first order equation in the far region is
%
\begin{eqnarray}
\frac{ d^{2}\eta^{(1)}}{dr^{2}} +\frac{n+1}{r}\frac{d \eta^{(1)}}{dr} -\lp \frac{n+1}{r^2}+n\hat{k}^{2}+\Omega^{2}\rp\eta^{(1)}= 
-\lp P^{(1)}\frac{d \eta^{(0)}}{dr} +Q^{(1)}\eta^{(0)}\rp,
\end{eqnarray} 
%
where
%
\beq
P^{(1)}=\frac{n(2n\hat k^{2}+3\Omega^{2})}{r^{n+1}k^{2}_{\Omega}}, \qquad
Q^{(1)}=\frac{n^2\hat k^{2}-(n^2\hat k^{4}+3n\hat k^{2}\Omega^{2}+2\Omega^{4})r^{2}}{r^{n+2}k^{2}_{\Omega}}. 
\eeq
%
Green's method gives the solution as
%
\begin{eqnarray}
\eta^{(1)}\,=\,A^{(1)}\frac{K_{\nu}(k_\Omega r)}{r^{n/2}} +S_{1}\frac{I_{\nu}(k_{\Omega} r)}{r^{n/2}}
+T_{1}\frac{K_{\nu}(k_{\Omega} r)}{r^{n/2}},
\end{eqnarray}
%
where
%
\beq
S_{1}=\int^{\infty}_{k_{\Omega}r} K_{\nu}(x)\mathcal{S}(x)dx\,, \qquad
T_{1}=\int^{k_{\Omega}r}_{k_{\Omega}\tilde r} I_{\nu}(x)\mathcal{S}(x)dx\,.
\eeq
%
$\mathcal{S}$ is a source term defined as
%
\begin{eqnarray}
\mathcal{S}\,=\,-k^{-1}_{\Omega}r^{\nu}\left( P^{(1)}\frac{d \eta^{(0)}}{dr} +Q^{(1)}\eta^{(0)} \right).
\end{eqnarray}
%
We can take $\tilde r$ arbitrarily since it does not affect the matching procedure. 

A very lengthy calculation yields $S_{1}$ and $T_{1}$ in an unilluminating form. For the purposes of matching one only needs the expressions in the overlap region. With the normalization of the leading order solution \eqref{farsoln} fixed like in sec.~\ref{subsec:farGL} (eq.~\eqref{farOL}), namely
\beq
\eta^{(0)}=A_n^{-1}\frac{K_\nu (k_\Omega r)}{r^{n/2}}\,,
\eeq
then the solution in this region is
%
\begin{eqnarray}
A_n\eta^{(1)}&=&\frac{1}{\sR^{2}} +\frac{\Omega^{2}-\hat{k}^{2}+\hat{k}^{4}-2\hat{k}^{2}(1+\hat{k}^{2})\ln{\sR}}{2n\hat{k}^{2}\sR^{2}} \notag \\
&&
+\frac{1}{2\hat{k}^{4}n^{2}\sR^{2}}\Bigl( \hat{k}^{2}\Omega^{2}-\Omega^{4}+\hat{k}^{4}+2\hat{k}^{4}\Omega^{2}
+3\hat{k}^{6}-\hat{k}^{8} \notag \\
&&\qquad\quad
+(\hat{k}^{4}-\Omega^{2}-3\hat{k}^{4}\Omega^{2}+2\hat{k}^{6}+\hat{k}^{8})\ln{\sR}
+\hat{k}^{4}(1+\hat{k}^{4})(\ln{\sR})^{2} \Bigr) \notag \\
&&
+O(n^{-3}). \label{farfirst}
\end{eqnarray}
%
The actual expression for $A_n$ is rather complicated (dependent on $n$, $\hat k$ and $\Omega$) and we omit it.

This solution yields terms $\propto \sR^{-2}$ at orders $n^0$, $n^{-1}$ and $n^{-2}$. We can match them to the corresponding terms in the large $\sR$ expansion of the near solutions $\eta_{(0)}$ \eqref{eta0}, $\eta_{(1)}$ \eqref{eta1}, and $\eta_{(2)}$ \eqref{eta2}. The first of these is just a matching of the overall amplitude. The second one and the third one would provide, respectively, the dispersion relations \eqref{Omvsk0} and \eqref{Omvsk1}.

\subsection{Sources for near-region equation}\label{app:sources}

The source term of the $n$-th order equation is given by
%
\begin{eqnarray}
\mathcal{S}_{(n)}=\sum_{j=1}^{n}S_{(j)}(\eta_{(n-j)}), 
\end{eqnarray}
%
where $\eta_{(j)}$ is the $j$-th order solution. The $S_{(j)}$ at each order are
%
\begin{eqnarray}
&&
S_{(1)}(\eta) = -2(\sR-1)^{3}\big[ 1+2\sR(-1+2\sR)\hat{k}^{2}\big]\eta' \notag \\
&&~~~~~~~~~~~~~~~
-(\sR-1)^{2}\big[3+(1-4\sR+8\sR^{2})\hat{k}^{2}\big]\eta, 
\end{eqnarray}
\begin{eqnarray}
&&
S_{(2)}(\eta) =
-8(\sR-1)^{3}\sR \big[ \sR\Omega^{2}+(-1+\sR+(-1+2\sR)\ln{\sR})\hat{k}^{2} \notag \\
&&~~~~~~~~~~~~~~~~~~~~~~~~~~~~~~~~~~~
+2\sR(1-3\sR+2\sR^{2})\hat{k}^{4} \big]\eta' \notag \\
&&~~~~~~~~~~~~
-\frac{\sR-1}{\sR}\big[ 1-2\sR +\sR^{2}(1-5\Omega^{2})-4\Omega^{2}\sR^{3} +8\Omega^{2}\sR^{4} \notag \\
&&~~~~~~~~~~~~
+2\sR(\sR-1)(4-12\sR+8\sR^{2}+(1-4\sR+8\sR^{2})\ln{\sR})\hat{k}^{2} \notag \\
&&~~~~~~~~~~~~~~~~~
+8\sR^{2}(\sR-1)^{2}(1-2\sR+4\sR^{2})\hat{k}^{4} \big] \eta,
\end{eqnarray}
\begin{eqnarray}
&&
S_{(3)}(\eta)=
-8(\sR-1)^{3}\sR\Big[ \Omega^{2} \sR(1+2\ln{\sR})+\hat{k}^{2}(2\sR^{2}(-3+4\sR)\Omega^{2} \notag \\
&&~~~~~~~~~~~~~~~~~~~
+2(-1+\sR)\ln{\sR} +(-1+2\sR)(\ln{\sR})^{2} \notag \\
&&~~~~~~
+4\sR(\sR-1)\hat{k}^{2}(-1+\sR+(-2+4\sR)\ln{\sR} +2\sR(\sR-1)(-1+2\sR)\hat{k}^{2}))  \Big] \eta' \notag \\
&&~~~
-2(\sR-1)\Big[ \sR\Omega^{2}(2(\sR-1)(-1+4\sR)+(-5-4\sR+8\sR^{2})\ln{\sR}) \notag \\
&&~~~~~~
+(\sR-1)\hat{k}^{2}(4(1+\sR(-2+\sR+\sR(-1-4\sR+8\sR^{2})\Omega^{2})) \notag \\
&&~~~~~~
+8(\sR-1)(-1+2\sR)\ln{\sR} +(1-4\sR+8\sR^{2})(\ln{\sR})^{2} \notag \\
&&~~~~~~
+16\sR(\sR-1)\hat{k}^{2}(1-3\sR+2\sR^{2}+(1-2\sR+4\sR^{2})\ln{\sR} \notag \\
&&~~~~~~~~~~
+\sR(\sR-1)(1-2\sR+4\sR^{2})\hat{k}^{2}) )\Big] \eta 
\end{eqnarray}
and
\begin{eqnarray}
&&
S_{(4)}(\eta)=-\frac{16\sR(\sR-1)^{3}}{3}\Big[ 
3\sR\Omega^{2}(2\sR^{2}\Omega^{2}+\ln{\sR}+(\ln{\sR})^{2})+\hat{k}^{2}(12\Omega^{2}\sR^{2}(\sR-1) \notag \\
&&~~~~~~~
+12\sR^{2}(-3+4\sR)\Omega^{2}\ln{\sR} +3(\sR-1)(\ln{\sR})^{2} +(-1+2\sR)(\ln{\sR})^{3} \notag \\
&&~~~~~~~~
+24(\sR-1)\sR\hat{k}^{2}(\sR^{2}(-2+3\sR)\Omega^{2}+(-1+\sR+(-1+2\sR)\ln{\sR})\ln{\sR} \notag \\
&&~~~~~~~~~
+(\sR-1)\sR\hat{k}^{2}(-1+\sR+(-3+6\sR)\ln{\sR}+2(\sR-1)\sR(-1+2\sR)\hat{k}^{2}))) \Big] \eta' \notag \\
&&
-\frac{2(\sR-1)}{3}\Big[ 3\sR\Omega^{2}(4(\sR-1)(-1+\sR+2\sR^{2}(1+2\sR)\Omega^{2}) +4(\sR-1)(-1+4\sR)\ln{\sR} \notag \\
&&~~~~~
+(-5-4\sR+8\sR^{2})(\ln{\sR})^{2})+2(\sR-1)\hat{k}^{2} (12(\sR-1)\sR^{2}(-3+8\sR)\Omega^{2} \notag \\
&&~~~~~~
+12(1+\sR(-2+\sR+2\sR(-1-4\sR+8\sR^{2})\Omega^{2})\ln{\sR} \notag \\
&&~~~~~~~~
+12(\sR-1)(-1+2\sR)(\ln{\sR})^{2} +(1-4\sR+8\sR^{2})(\ln{\sR})^{3} \notag \\
&&~~~~~~~~~~
+24(\sR-1)\sR\hat{k}^{2}((\sR-1)^{2}+6\sR^{3}(-1+2\sR)\Omega^{2} \notag \\
&&~~~~~~~~~~~
+2\ln{\sR}(2-6\sR+4\sR^{2}+(1-2\sR+4\sR^{2})\ln{\sR}) \notag \\
&&~~~~~~~~~~~
+2(\sR-1)\sR\hat{k}^{2}(2-6\sR+4\sR^{2}+3(1-2\sR+4\sR^{2})\ln{\sR} \notag \\
&&~~~~~~~~~~~~~
+2(\sR-1)\sR(1-2\sR+4\sR^{2})\hat{k}^{2}))) \Big] \eta.
\end{eqnarray}
%

\paragraph{Sources at large $\sR$.}

The integrations constants $A_{i}$ and $B_{i}$ at $i$-th order in \eqref{greenfn}
 are determined by the boundary conditions at $\sR=1$ and $\sR\gg 1$. This $i$-th order solution contributes to the large $\sR$ condition for the $(i+1)$-th order 
solution through $S_{(1)}(\eta_{(i)})$. Since $S_{(i)}(\eta)$ is linear in
$\eta$, the contributions of $A_{i}$ and $B_{i}$ can be obtained independently. Now,
%
\begin{eqnarray}
S_{(1)}(A_{i}u_{0} +B_{i}v_{0})&=&-A_{i}(1+\hat{k}^{2})\sR-B_{i}(8\sR^{2}-12\sR)\hat{k}^{2}+O(\sR^{0}),
\end{eqnarray}
%
so the leading term at large $\sR$ is controlled by $B_i$, not $A_i$.
This source term is contained in the condition at large $\sR$ as
%
\begin{eqnarray}
\frac{1}{\sR^{2}}\int^{\sR}u_{0}(\sR')S_{(1)}(A_{i}u_{0} +B_{i}v_{0})d\sR' = -4\hat{k}^{2}B_{i}+O(\sR^{-1})\,, 
\end{eqnarray}
%
\ie\ only $B_{i}$ enters in the large $\sR$ condition for the $(i+1)$-th order solution. This is  why in sec.~\ref{subsec:nearreg} we can proceed from the third to the fourth order without $A_{3}$.

\section{Hydrodynamic vs.\ large $D$ expansions}\label{app:hydrovsn}

When refs.~\cite{Camps:2010br,Caldarelli:2012hy} compared the hydrodynamical and numerical calculations of $\Omega(k)$, it was observed that the agreement improves for larger $n$. The proposed explanation for this relies on a conjecture about the large $n$-dependence of higher-order hydrodynamic transport coefficients. More precisely, the large $D$ expansion in this article is of the form
\beq
\Omega=\hat k \lp 1-\hat k +\frac{b_1(\hat k)}{n}+\frac{b_2(\hat k)}{n^2}+\dots\rp\,,
\eeq
In the hydrodynamic expansion it seems more appropriate to keep fixed $T = n/(4\pi r_0)$, instead of $r_0$. Rescaling $\tilde\Omega=(n+1)\Omega$, $\tilde k=\sqrt{n+1}\,k$, the expansion is 
\beqa\label{hydro}
\tilde\Omega=\tilde{k}\lp 1-\frac{n+2}{n+1}\frac{\tilde k}{4\pi T}+c_1(n)\lp\frac{\tilde k}{4\pi T}\rp^2+c_2(n)\lp\frac{\tilde k}{4\pi T}\rp^3+ \dots\rp\,.
\eeqa
The expansion variables in the two cases are equivalent at large $n$, since then $\hat k=\tilde k/(4\pi T)$. Nevertheless, these are clearly different expansions. It is apparent that the large-$n$ accuracy of the hydrodynamic results requires that $c_j(n)\to 0$ when $n\to\infty$ for all $j$ \cite{Camps:2010br}. Hence, lacking an explicit calculation of, say, $c_2(n)$, we could not be sure within the hydrodynamic approach (\ie\ prior to comparing to the numerical results) that at large $n$ there will not remain a term $\propto \hat k^4$ in the dispersion relation at leading order.

The $c_j(n)$ are obtained from effective hydrodynamic transport coefficients computed from black brane perturbations. While it seems possible that their leading large $n$ scaling can be determined from generic features of higher-dimensional gravity, the required behavior has not been derived yet within the hydrodynamic approach.
Our calculation of the dispersion relation in the large $n$ expansion can be regarded as a proof of it.

The hydrodynamic expansion applies also for non-linear black brane perturbations. It should be interesting to investigate if the large $D$ expansion can be useful in that problem too.


\end{document}